\documentclass[aps,prc,twocolumn,amsmath,superscriptaddress,floatfix,nofootinbib]{revtex4-1}

\usepackage{setspace,ulem, epsfig,amssymb,amsfonts,amsmath,mathtools,bm,color,xcolor,graphicx,braket,adjustbox,esint,upgreek,verbatim,ctable}
\usepackage[bookmarksnumbered,bookmarksopen,colorlinks,citecolor=blue,linkcolor=red]{hyperref}
\usepackage{graphicx}  
\usepackage{booktabs}

\usepackage{float}
 
\usepackage{graphicx}
\graphicspath{
    {./figs-1/}  
    {./fig/} 
}

\begin{document}
\title{
Nuclear Deformation Effects on Charmonium Suppression in Au+Au and U+U Collisions}

\author{Jiamin Liu}
\affiliation{Department of Physics, Tianjin University, Tianjin 300350, China}

\author{Huanshang Yang}
\affiliation{College of Engineering, Georgia Institute of Technology, Atlanta, GA 30332, United States}

\author{Baoyi Chen}
\email[]{baoyi.chen@tju.edu.cn}
\affiliation{Department of Physics, Tianjin University, Tianjin 300350, China}
\affiliation{Tianjin University-National University of Singapore Joint Institute in Fuzhou, Fuzhou 350207, China
}

\date{\today}
\begin{abstract}
We investigate the impact of intrinsic nuclear deformation and orientation on the yield suppression and momentum anisotropy of charmonia in Au+Au and U+U collisions at the Relativistic Heavy-Ion Collider (RHIC). The anisotropic nucleon density within the nucleus is parameterized using a modified Woods-Saxon distribution, which is incorporated into the initial distributions of both the heavy quarkonia and the bulk medium energy density. The well-established Boltzmann-type transport equation is utilized to describe the dynamical evolutions of quarkonium in the anisotropic bulk medium.  Treating quarkonium suppression in Au+Au collisions as a baseline, we find that the momentum-integrated charmonium yield suppression is relatively insensitive to the initial nuclear geometry in deformed U+U collisions. In contrast, the anisotropic flow coefficients ($v_n$) of the charmonium is more sensitive to the nuclear deformation. Furthermore, these observables are also connected with the collision configuration, particularly when distinguishing between tip–tip and body–body orientations in U+U collisions at $\sqrt{s_{NN}} = 193$ GeV. This effect is more pronounced for the excited state, $\psi(2S)$, due to its smaller binding energy and heightened sensitivity to the initial energy density of the hot QCD medium.
\end{abstract}

\maketitle

\section{Introduction} 
A hot deconfined matter is believed to be generated in relativistic heavy-ion collisions, which turns out to be nearly a perfect fluid~\cite{Bazavov:2012PRD,Heinz:2013th}.  
A realistic description of the nuclear density which incorporates the nuclear deformation is necessary for the studies of both light and heavy hadrons. 
The STAR Collaboration demonstrated that intrinsic nuclear shapes, such as quadrupole deformation~\cite{Giacalone:2020awm}, are intimately linked to the final-state momentum anisotropies of light hadrons~\cite{STAR:2024wgy}. In isobar collisions, nuclear deformation and neutron-skin effects can be disentangled using precision flow observables, indicating that distinct nuclear-structure ingredients leave identifiable imprints on the flow coefficients $v_n$~\cite{Jia:2022qgl}. Furthermore, orientation-dependent collisions, such as tip–tip versus body–body configurations, have also been investigated in U+U collisions. In these collisions, the initial energy densities of the bulk medium and their corresponding final-state momentum distributions of light hadrons are distinctly resolvable~\cite{Nepali:2007an,Goldschmidt:2015kpa}.

As heavy quarks and quarkonia are produced prior to the formation of the deconfined medium, they are expected to be uniquely sensitive to the initial geometric shape of the bulk energy density~\cite{Andronic:2015wma,Prino:2016cni}. Throughout their evolution, these heavy probes carry imprints of the anisotropic medium effects along their distinct trajectories~\cite{STAR:2017kkh}. While the anomalous suppression of charmonium was proposed four decades ago as a clear signature of the hot deconfined medium~\cite{Matsui:1986dk}, we extend this concept in the present work. Specifically, we propose utilizing the final momentum anisotropy of heavy quarkonia as a probe of the initial bulk geometry originating from nuclear deformation. Quarkonium dynamics in nuclear collisions are investigated herein using a Boltzmann-type transport model, which has successfully described the nuclear modification factors ($R_{AA}$) for both charmonium and bottomonium observables across a wide range of energies, from RHIC to the LHC~\cite{Zhu:2004nw,Yan:2006ve,Liu:2009nb,Zhou:2014kka,Chen:2016dke,Zhao:2022ggw}. It has also been employed to study the triangular flow ($v_3$) of $J/\psi$ in Pb+Pb collisions~\cite{Zhao:2021voa}, which is driven by fluctuations in the initial medium energy density. Motivated by these findings, one expects that nuclear deformation will similarly influence the initial anisotropic distribution of the QGP, thereby manifesting in the final-state $v_n$ of charmonia. 

In this work, we implement a deformed Woods-Saxon nuclear distribution with a tunable triaxiality angle $\gamma$ within an optical Glauber model~\cite{Miller:2007ri,Bally:2021qys} to generate geometry-dependent initial conditions of the bulk medium, which are further evolved with the viscous hydrodynamic model. We utilize the transport model to calculate and analyze the nuclear modification factors $R_{AA}$ and the momentum anisotropic flow coefficients $v_n$ for both $J/\psi$ and $\psi(2S)$ in the bulk medium.

\section{Theoretical framework}
\label{sec:theory}

\subsection{Hydrodynamic and transport models}

\begin{figure}[htp]
    \centering
    \includegraphics[scale=0.35]{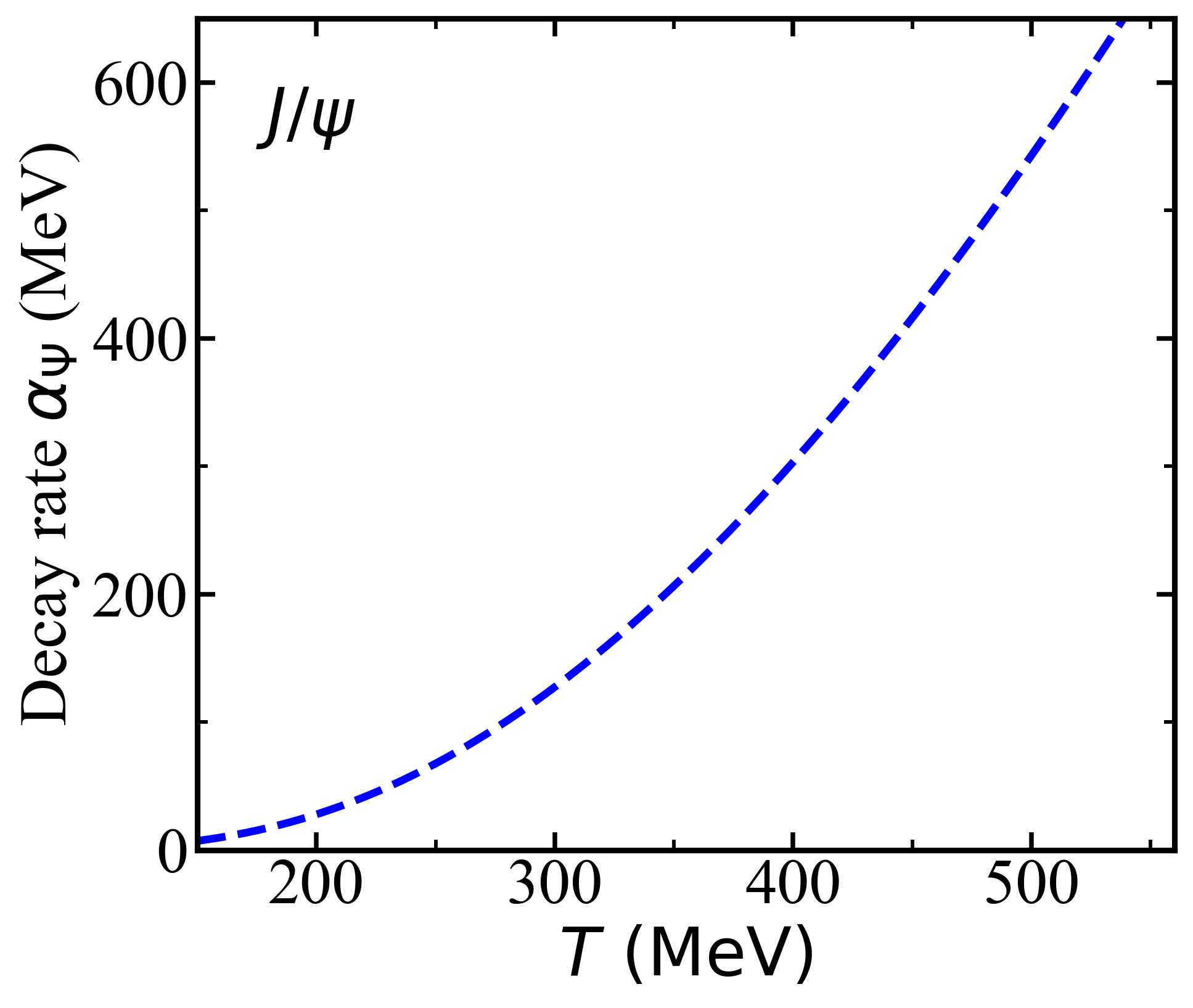}
    \caption{$J/\psi$ decay rate $\alpha_{J/\psi}(T)$ attributed to the inelastic scatterings with thermal gluons as a function of temperatures, the transverse momentum of $J/\psi$ is set to be $p_T=0$ used in Eq.(\ref{lab-decayrate}). }
    \label{fig:decay-rate}
\end{figure}

We employ the viscous hydrodynamic model, MUSIC package~\cite{Schenke:2010nt}, to simulate the spacetime evolution of the hot QCD medium. Regarding the Equation of State (EoS) used in the hydrodynamic model, we utilize a lattice QCD-based parametrization for the deconfined phase, matched to a Hadron Resonance Gas model for the hadronic phase~\cite{Huovinen:2009yb}. Within this background, the phase-space evolution of heavy quarkonia is governed by the Boltzmann transport equation~\cite{Zhu:2004nw,Zhou:2014kka,Chen:2018kfo},
\begin{align}
\label{eq-tra}
&\Bigg[
  \cosh(y-\eta)\,\frac{\partial}{\partial\tau}
  + \frac{\sinh(y-\eta)}{\tau}\,\frac{\partial}{\partial\eta}
  + \mathbf{v}_T\cdot\nabla_T
\Bigg] f_\Psi  \nonumber \\
&= -\alpha_\Psi\, f_\Psi + \beta_\Psi,
\end{align}
with the definition of the proper time $\tau = \sqrt{t^2 - z^2}$, the particle rapidity$y = \frac{1}{2} \ln[(E + p_z)/(E - p_z)]$, and the pseudo rapidity $\eta = \frac{1}{2} \ln[(t + z)/(t - z)]$. The spatial diffusion of charmonia is captured by the three terms on the left-hand side (L.H.S.) of Eq.~(\ref{eq-tra}), where $\mathbf{v}_T = \mathbf{p}_T/E_T$ denotes the transverse velocity and $E_T = \sqrt{m_\Psi^2 + p_T^2}$ is the transverse energy. Here, $m_\Psi$ represents the vacuum mass of the charmonium eigenstate. Within the QGP, charmonium dissociation is driven by inelastic collisions with thermal partons, which is incorporated into the transport equation via the decay rate $\alpha_\Psi$. The magnitude of this decay rate~\cite{Zhu:2004nw,Chen:2015iga,Gerschel:1988wn} depends on both the local medium temperature and the inelastic cross section,
\begin{equation}
\label{lab-decayrate}
\alpha_\Psi
=
\begin{aligned}[t]
  &\frac{1}{2E_T} \int \frac{d^3\mathbf{k}}{(2\pi)^3\,E_g}\,
    \sigma_{g\Psi}(\mathbf{p},\mathbf{k},T)\,
  \\
  &\qquad\times 4F_{g\Psi}(\mathbf{p},\mathbf{k})\,
    f_g(\mathbf{k};T)\,,
\end{aligned}
\end{equation}
where the variables $\boldsymbol{k}$ and $E_g$ denote the momentum and energy of the gluon, respectively. The gluon thermal distribution, $f_g$, is assumed to follow a massless Bose-Einstein distribution. $F_{g\Psi}$ represents the flux factor for the gluo-dissociation process $g+\Psi\rightarrow c+\bar{c}$. The cross section $\sigma_{g\Psi}$ for $J/\psi$~\cite{Zhu:2004nw} is calculated using the Operator Product Expansion (OPE) method~\cite{PESKIN1979365,BHANOT1979391}, while the cross sections for excited charmonium states are obtained via geometric scaling: $\sigma_{g\psi(2S)} = \sigma_{gJ/\psi} \times \langle r_{\psi(2S)} \rangle^2 / \langle r_{J/\psi}\rangle^2$~\cite{Chen:2018kfo}. The $J/\psi$ decay rate calculated with Eq.(\ref{lab-decayrate}) is shown in Fig.~\ref{fig:decay-rate}. In nuclear collisions where multiple charm-quark pairs are produced in the deconfined medium, both theoretical and experimental studies indicate that final-state charmonium production can be enhanced through the coalescence of charm and anti-charm quarks~\cite{Braun-Munzinger:2000csl,Du:2015wha,Blaizot:2017ypk,Chen:2021akx}. This process primarily contributes to the low-$p_T$ region. However, since this study focuses on probing the nuclear deformation of Uranium ($U$) at RHIC energies using high-$p_T$ charmonium, the coalescence process is not the dominant production mechanism~\cite{Zhao:2010nk}. Consequently, the regeneration term $\beta_\Psi$ can be neglected in the following calculations.

\subsection{Initial momentum distribution}
\label{subsec:initial_mom}

To solve the transport equation, the initial charmonium distribution is needed. The momentum distribution of $J/\psi$ in $AA$ collisions, $d^2 N_\Psi / d y d p_T$, is parametrized as the product of the rapidity-differential cross section, $d \sigma_\Psi / d y$, and a normalized $p_T$ distribution $dN_\Psi/dp_T$. Note that in the calculation of $R_{AA}$ and $v_n$, the overall normalization provided by $d \sigma_\Psi / d y$ cancels out. The $p_T$ dependence can be parametrized with the form according to the experimental data~\cite{PHENIX:2006aub}:
\begin{align}
  \frac{dN_\Psi}{dp_T}
  \propto
  p_T
  \left[
    1+\frac{p_T^2}{(n-2)\,\langle p_T^2\rangle(y)}
  \right]^{-n}.
  \label{eq:tsallis}
\end{align}
The rapidity dependence of the mean-squared transverse momentum is parametrized as
\begin{equation}
  \langle p_T^2\rangle(y)
  = \langle p_T^2\rangle(y=0)\,
    \Big[1-\big(y/y_{\max}\big)^2\Big].
\end{equation}
where the maximum rapidity is defined as $y_{\max} = \operatorname{arccosh}(\sqrt{s_{NN}} / 2m_{J/\psi})$. Based on the $J/\psi$ distribution in $pp$ collisions at RHIC energies, we set the parameters to $n=6$ and $\langle p_T^2\rangle_{pp}(y=0)=4.14\ \rm{(GeV/c)^2}$~\cite{Liu:2009wza} for $J/\psi$ at $\sqrt{s_{NN}} = 200$ and $193$ GeV. The initial spatial distribution of charmonia is assumed to be proportional to the product of the nuclear thickness functions, i.e., $dN_\Psi/d\mathbf{x}_T \propto T_A(\mathbf{x}_T + \mathbf{b}/2) T_B(\mathbf{x}_T - \mathbf{b}/2)$. The thickness function is defined as the longitudinal integral of the nucleon density, $T(\mathbf{x}_T) = \int dz \rho(\mathbf{x}_T, z)$. Information regarding nuclear deformation is encoded within the density distribution $\rho(\mathbf{x}_T, z)$, the details of which are presented in the following section.

\subsection{Deformed nuclear density}
\label{sec:init_geometry}

\begin{figure}[htp]
  \centering
\includegraphics[width=0.8\linewidth]{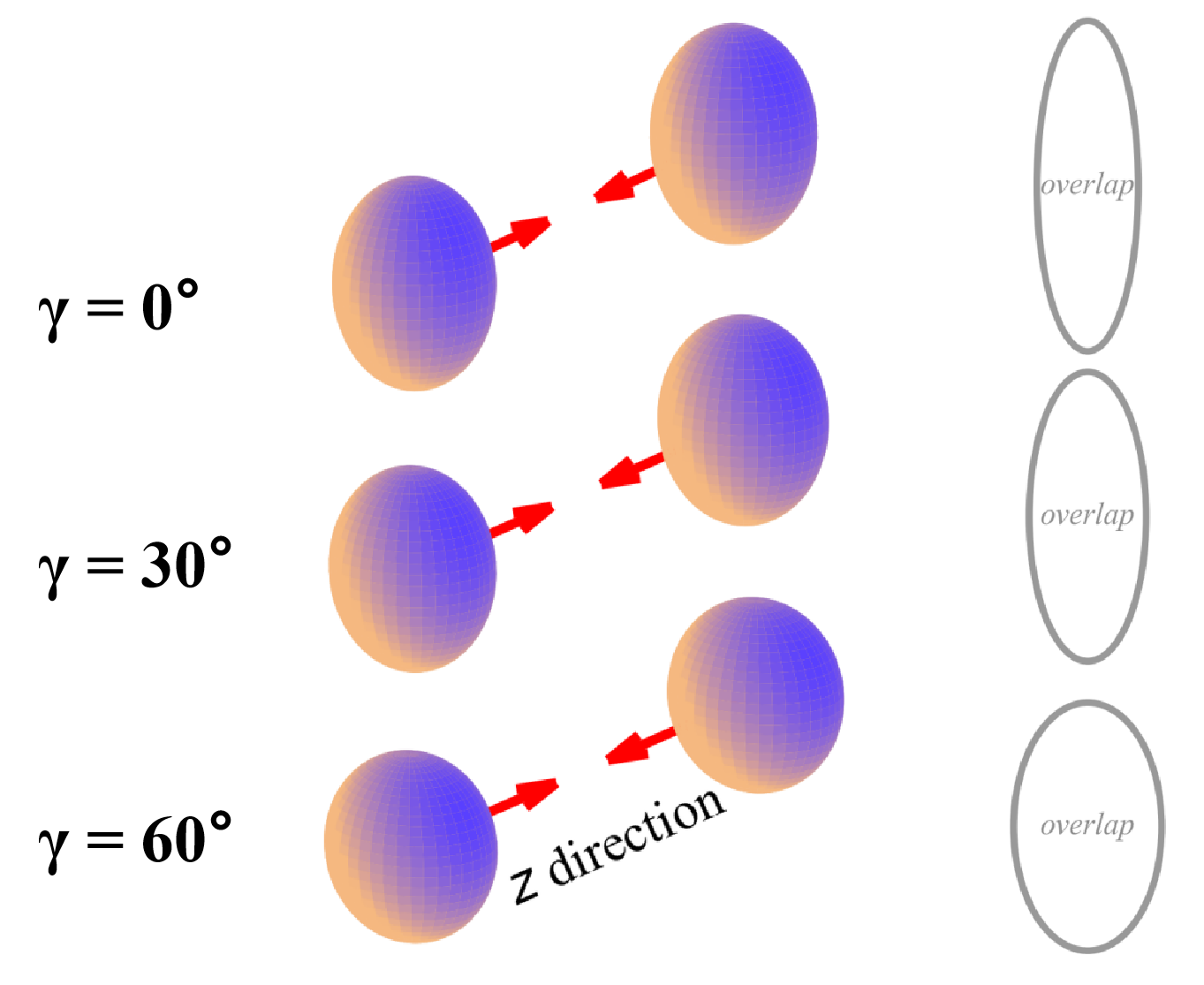}
\caption{
Schematic illustration of the impact of the triaxiality angle $\gamma$ in Eq.~(\ref{eq:R_deformed}) on the initial geometry of the overlap between two nuclei.
Ellipsoids denote the deformed nuclei; red arrows indicate the beam ($z$-) direction.
The gray contours on the right show the transverse overlap projection for $\gamma=0^\circ,\,30^\circ,\,60^\circ$.}
\label{fig:gamma-collision}
\end{figure}

\begin{figure}[htp]
  \centering
\includegraphics[width=0.8\linewidth]{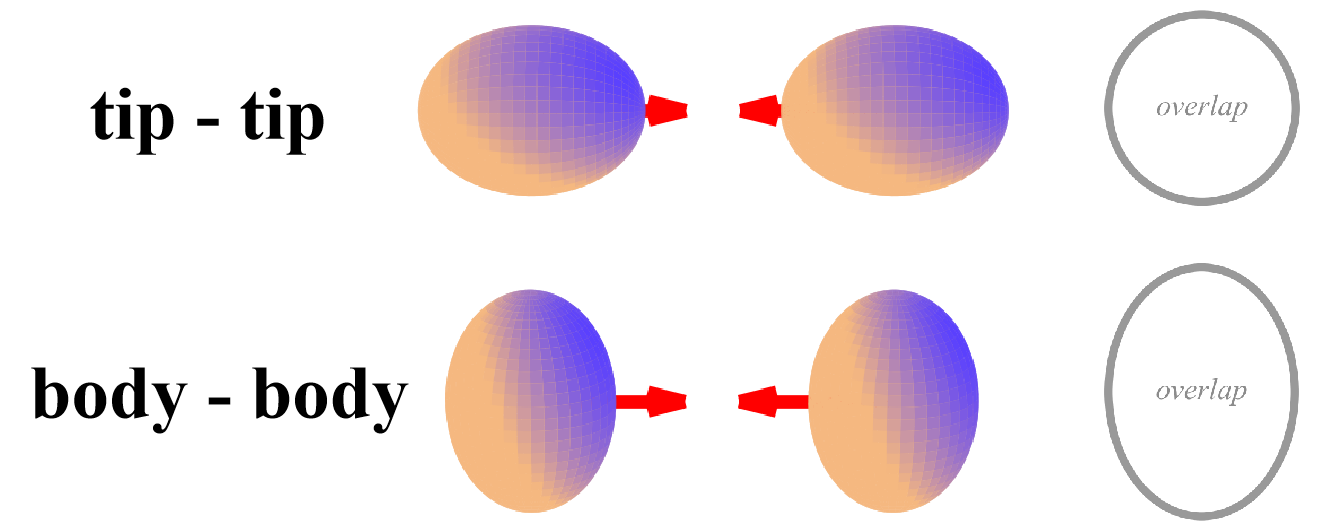}
\caption{
Definition of the idealized tip-tip and body-body configurations at $\gamma=0^\circ$.
Tip-tip: the long axes of both nuclei are aligned with the beam ($z$) direction; body-body: the long axes lie in the transverse plane.
The gray contours on the right show the corresponding transverse overlap projections.
}
\label{fig:tipbody-collision}
\end{figure}

The nuclear density, derived from the optical Glauber model and incorporating nuclear deformation effects, can be parametrized as follows~\cite{Lu:2023fqd}
\begin{equation}
  \rho_A(r',\theta',\phi';\gamma)
  = \frac{\rho_{0,A}}{
    1 + \exp\!\left[\dfrac{r' - R_A(\theta',\phi';\gamma)}{a_A}\right]} \,,
  \label{eq:ws}
\end{equation}
where $r'$, $\theta'$, and $\phi'$ denote the radial coordinate, polar angle, and azimuthal angle, respectively. The normalization constant $\rho_{0,A}$ is determined by the condition $\int d^3r'\,\rho_A = A$, with A to be the nucleon number. $R_A(\theta', \phi'; \gamma)$ characterizes the nuclear shape. For nuclei exhibiting quadrupole and hexadecapole deformations, this radius is parameterized as~\cite{STAR:2024wgy,Zhao:2024lpc,Mascalhusk:2024fjp,Schenke:2014tga},
\begin{align}
  R_A(\theta',\phi';\gamma)
  &= R_{0,A}\Big[
    1 + \beta_2\big(\cos\gamma\,Y_{20}(\theta',\phi')
     \nonumber\\
  &\quad
    + \sin\gamma\,Y_{22}^{(c)}(\theta',\phi')\big)
    + \beta_4 Y_{40}(\theta',\phi')
  \Big],
  \label{eq:R_deformed}
\end{align}
where the parameter $\gamma$ denotes the Bohr triaxiality angle, which interpolates between prolate-like and oblate-like quadrupole shapes by mixing the $m=0$ and $m=\pm2$ components of the $\ell=2$ deformation. We utilize the real tesseral combination:
$$Y_{22}^{(c)} \equiv (Y_{22} + Y_{2,-2})/\sqrt{2} = \sqrt{\frac{15}{16\pi}}\sin^2\theta'\cos(2\phi'),$$such that $\gamma=0^\circ$ corresponds to a purely axial (prolate) quadrupole deformation aligned with the intrinsic $z'$-axis, whereas $\gamma=60^\circ$ represents an oblate deformation. Unless stated otherwise, we select $\gamma=0^\circ, 30^\circ,$ and $60^\circ$ as representative cases for the orientation-averaged scans. The corresponding change in the transverse overlap geometry induced by varying $\gamma$ is illustrated in Fig.~\ref{fig:gamma-collision}.For the idealized tip–tip and body–body configurations discussed below, we restrict our analysis to $\gamma=0^\circ$ to provide unambiguous definitions of the long-axis orientation relative to the beam. The hexadecapole component, $\beta_4 Y_{40}$, is maintained as axial and independent of $\gamma$, following the conventions established in Ref.~\cite{Giacalone:2020awm,Jia:2021tzt}. The idealized tip-tip and body-body configurations and their transverse overlap projections are presented in Fig.~\ref{fig:tipbody-collision}.
The parameters for the deformed Woods-Saxon distributions of $^{197}$Au and $^{238}$U are summarized in Table~\ref{tab:nuclearparams}. For $^{197}$Au, we consider both spherical and deformed configuration with $(\beta_2, \beta_4) = (-0.13, -0.03)$. In contrast, $^{238}$U is treated as a strongly prolate nucleus with $(\beta_2, \beta_4) = (0.28, 0.093)$, consistent with previous studies at RHIC energies~\cite{Giacalone:2020awm,Jia:2021tzt}.

\begin{table}[!htp]
  \centering
  \caption{The values of the parameters in deformed Woods-Saxon distribution Eq.(\ref{eq:ws}). }
  \label{tab:nuclearparams}
  \begin{tabular}{lcccc}
    \hline\hline
    Nucleus & $R_{0}$ (fm) & $a_A$ (fm) & $\beta_2$ & $\beta_4$ \\
    \hline
    $^{238}$U & 6.81 & 0.55  & 0.28  & 0.093 \\
    $^{197}$Au (symmetric) & 6.42 & 0.41 & 0 & 0 \\
    $^{197}$Au (deformed)  & 6.42 & 0.41 & $-0.13$ & $-0.03$ \\
    \hline\hline
  \end{tabular}
\end{table}

In the theoretical calculations, an average impact parameter, $\langle b \rangle$, is employed to represent a specific collision centrality for simplicity. The mean impact parameter values for various centralities~\cite{PHENIX:2011img} are summarized in Table~\ref{tab:centrality}.

\begin{table}[!htp]
  \centering
  \caption{Centrality bins and representative impact parameters $\langle b\rangle$
  used for the comparison with experimental $J/\psi$ data in Au+Au collisions at $\sqrt{s_{NN}}=200$~GeV.}
  \label{tab:centrality}
  \begin{tabular}{ccc}
    \hline\hline
      & $\langle b\rangle$ (fm) & Centrality \\
    \hline
    $v_2$ bins & 2.2  & 0--10\% \\
               & 6.7  & 10--40\%  \\
               & 10.7 & 40--80\%  \\
               & 6.2  & 0--80\%   \\
    \hline
    $R_{AA}$ bins & 3.3   & 0--20\%   \\
                  & 7.9  & 20--40\%  \\
                  & 10.4 & 40--60\%  \\
                  & 5.8   & 0--60\%   \\
    \hline\hline
  \end{tabular}
\end{table}

For a given orientation of each nucleus relative to the beam axis, specified by the Euler angles $\Omega_A$, the nuclear thickness function is defined as:
\begin{equation}
T_{A}(x,y;\Omega_A,\gamma) = \int_{-\infty}^{+\infty} dz \rho_{A}\big(r'(x,y,z;\Omega_A),\theta',\phi';\gamma\big).
\label{eq:thickness}
\end{equation}
For nucleus B, the definition of the thickness function is the same. Within the optical Glauber model, the local density of binary nucleon–nucleon collisions at the transverse coordinate $\boldsymbol{x}_T=(x,y)$ and impact parameter $\mathbf{b}$ is given by:
\begin{equation}
  \begin{aligned}
  n_{\text{coll}}(\boldsymbol{x}_T;\mathbf{b},\Omega_A,\Omega_B,\gamma)
  &= \sigma_{NN}^{\text{inel}}
  \,T_A\!\left(\boldsymbol{x}_T + \tfrac{\mathbf{b}}{2};\Omega_A,\gamma\right) \\
  &\quad \times
  T_B\!\left(\boldsymbol{x}_T - \tfrac{\mathbf{b}}{2};\Omega_B,\gamma\right),
  \end{aligned}
  \label{eq:ncoll}
\end{equation}
where $\sigma_{NN}^{\text{inel}}$ denotes the nucleon–nucleon inelastic cross section. In the transport model, the binary collision density serves as the relative initial spatial distribution for primordially produced charmonium.

\section{Entropy-density profiles with deformed nucleus}
\label{subsec:profiles}

\begin{figure*}[!htp]
  \centering
\includegraphics[width=0.4\linewidth]{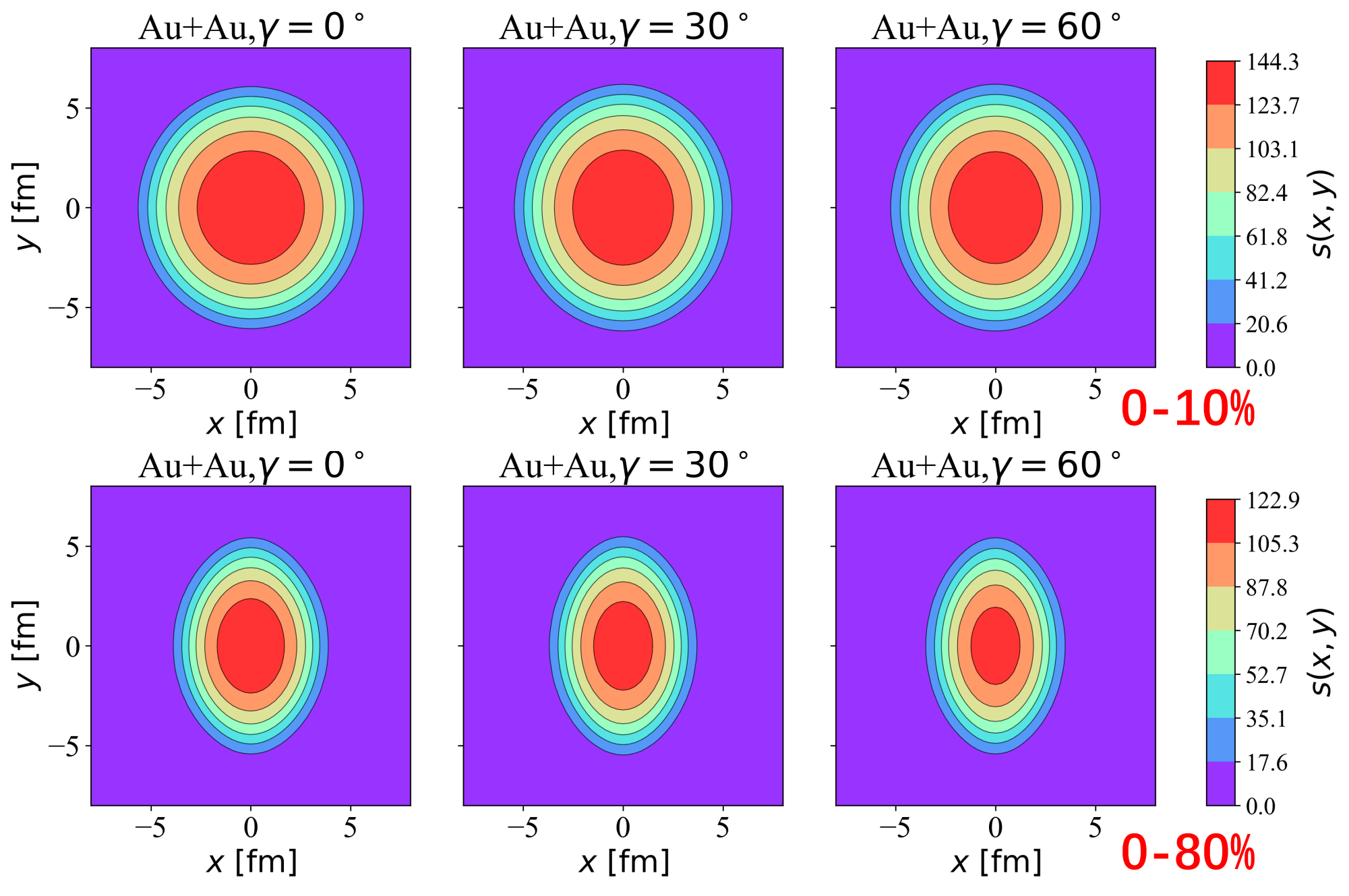}    
\includegraphics[width=0.4\linewidth]{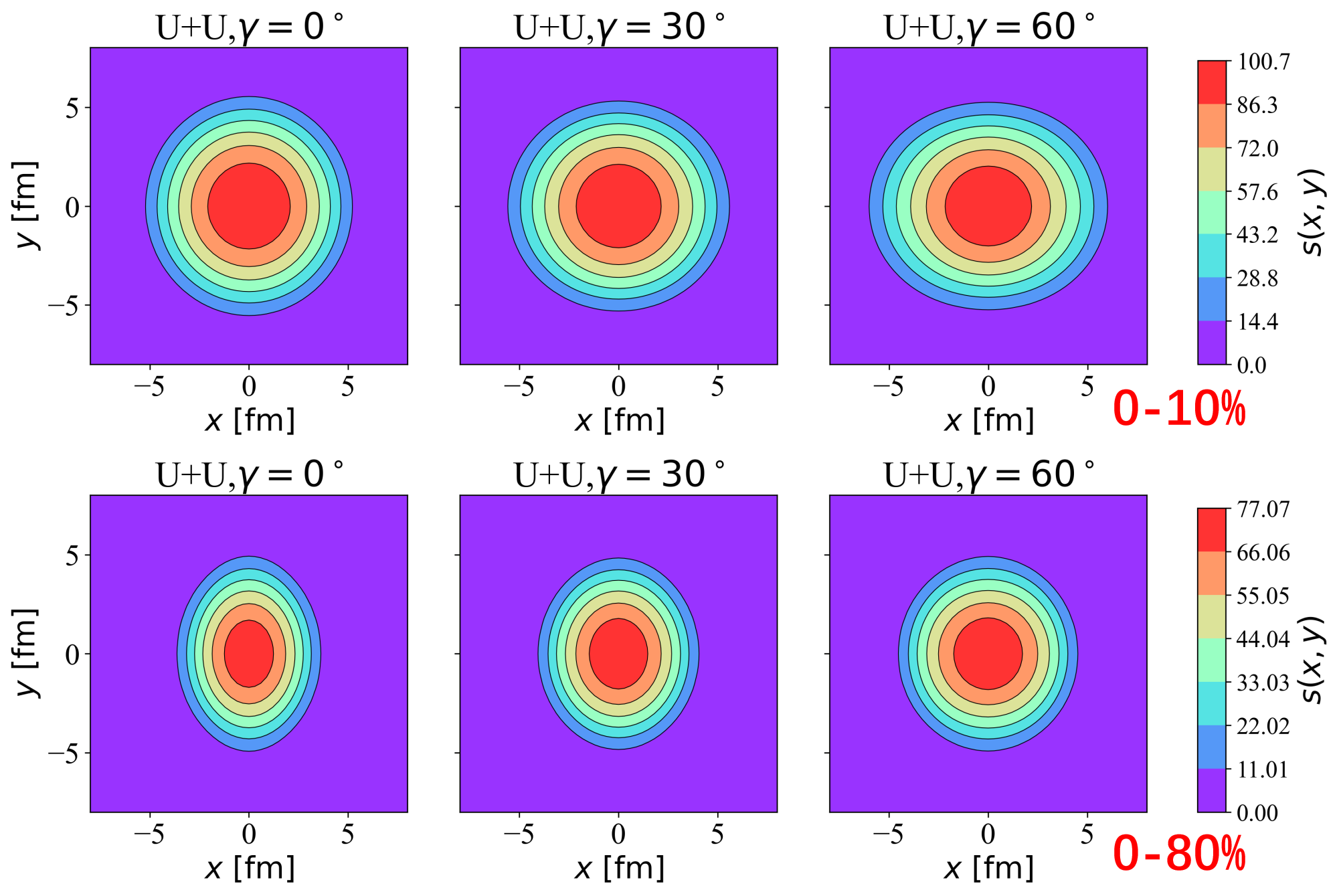}
  \caption{
Initial entropy density profiles $s_n(\tau_0, x, y)$ for 200 GeV deformed Au+Au (left), and 193 GeV deformed U+U (right) collisions at the initial time $\tau = \tau_0$. The sub-panels display collision centralities of 0–10\% and 0–80\%, respectively. The Bohr triaxiality angle $\gamma$ is set to $0^\circ, 30^\circ, \text{and } 60^\circ$ within the nuclear density distribution for Uranium ($U$).
  }
  \label{fig:entropy_profiles-GAMMA}
\end{figure*}

\begin{figure*}[!htp]
  \centering
\includegraphics[width=0.4\linewidth]{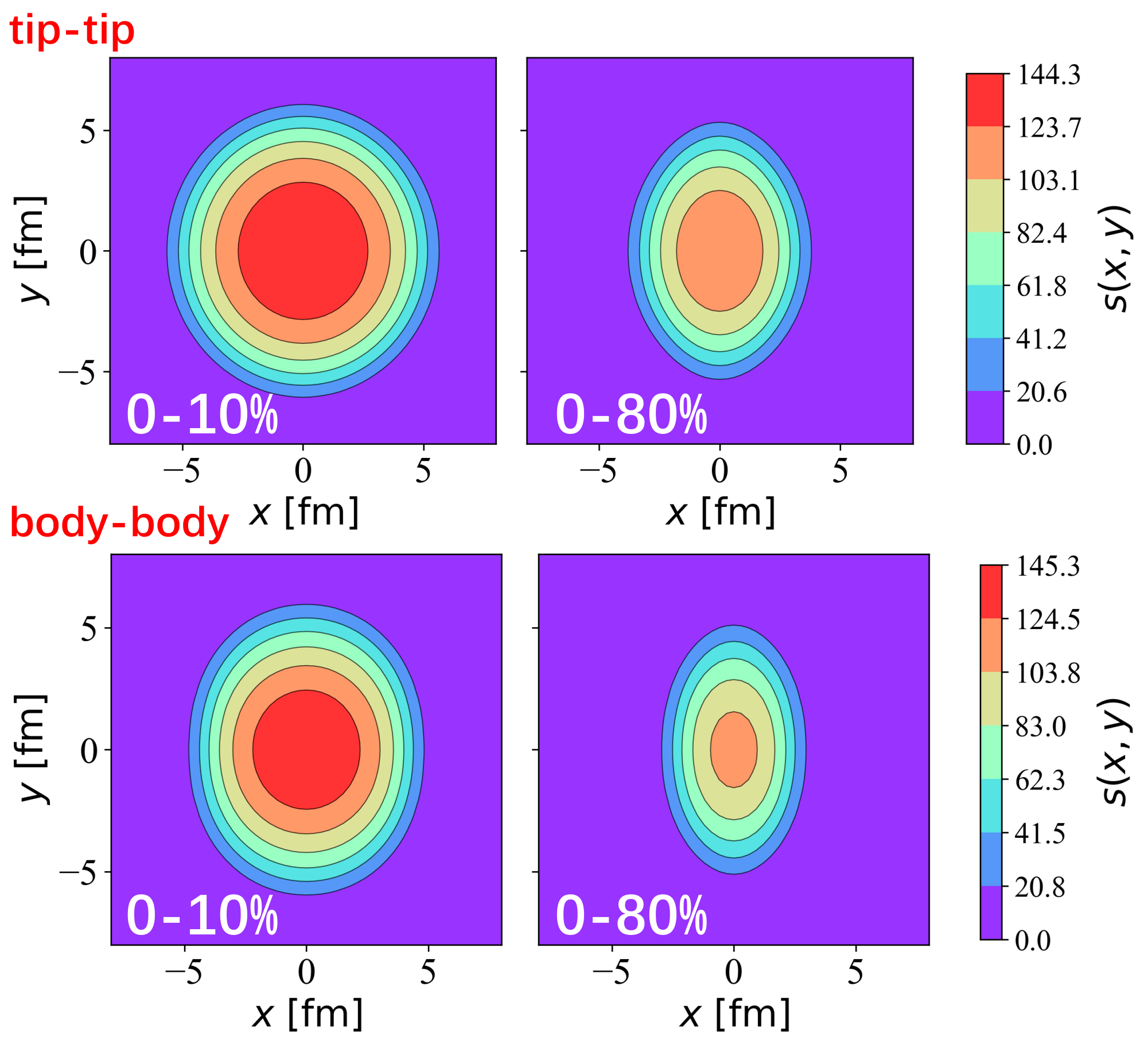}
\includegraphics[width=0.4\linewidth]{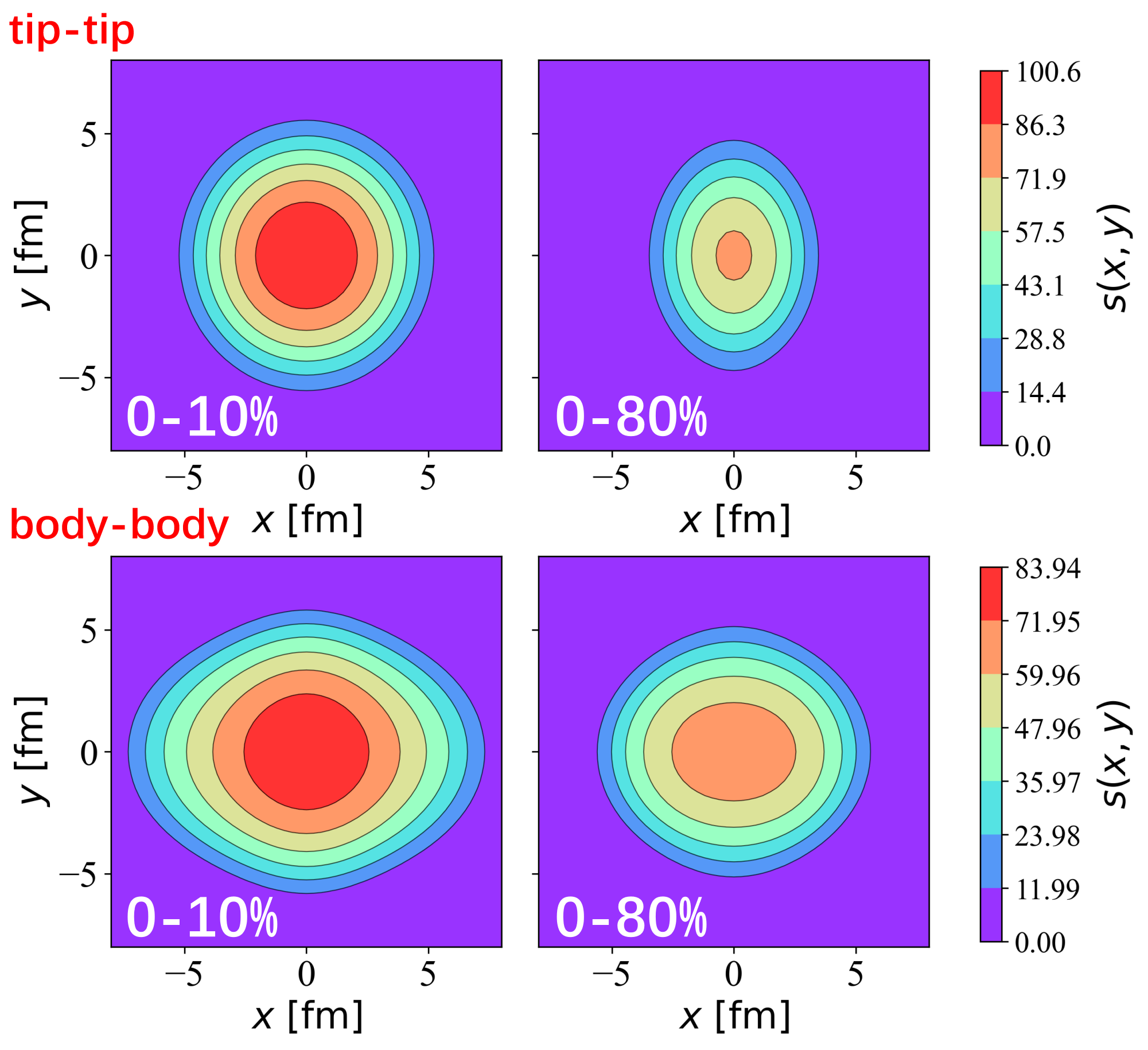}
  \caption{Initial entropy density profiles $s_n(\tau_0, x, y)$ for deformed Au+Au (left), and deformed U+U (right) collisions. Both tip-tip and body-body configurations with a Bohr triaxiality angle $\gamma = 0^\circ$ are displayed.
  }
  \label{fig:entropy_profiles-body}
\end{figure*}

The deformed nuclear density not only modifies the initial spatial distribution of heavy quarkonia but also the initial entropy density of the bulk medium. This initial eccentricity is subsequently converted into the final-state momentum anisotropy of light hadrons through hydrodynamic expansion. The Bohr triaxiality angle $\gamma$, defined in Eq.~(\ref{eq:R_deformed}), is set to $\gamma = 0^\circ, 30^\circ,$ and $60^\circ$ to systematically investigate its impact. These variations alter the nucleon density and the initial entropy density profiles, as illustrated in Fig.~\ref{fig:entropy_profiles-GAMMA}. In the left panel of Fig.\ref{fig:entropy_profiles-GAMMA}, the deformed Au+Au is plotted as a baseline. For U+U collisions, the deformed effect is a bit more pronounced as shown in the right panel of Fig.~\ref{fig:entropy_profiles-GAMMA}. 

To further disentangle the roles of extreme orientations, the initial entropy densities for tip–tip and body–body configurations in Au+Au  and U+U collisions are presented in left and right panels of Fig.~\ref{fig:entropy_profiles-body} respectively. Two centralities 0–10\% and 0–80\% are selected. The right panel is for U+U collisions. In the tip–tip geometry of U+U 0-10\% collisions, the long axes of both nuclei are aligned with the beam direction ($z$-axis), yielding a compact, nearly circular overlap region and relatively short transverse path lengths. Conversely, in the body–body geometry, the long axes lie within the transverse plane, producing a more extended and strongly elliptic fireball with a lower peak entropy density but substantially increased path lengths along the $x$-axis.
For deformed Au+Au collisions, the difference between the tip–tip and body–body profiles remains moderate. 

\begin{figure}[!htp]
  \centering
\includegraphics[width=0.42\linewidth]{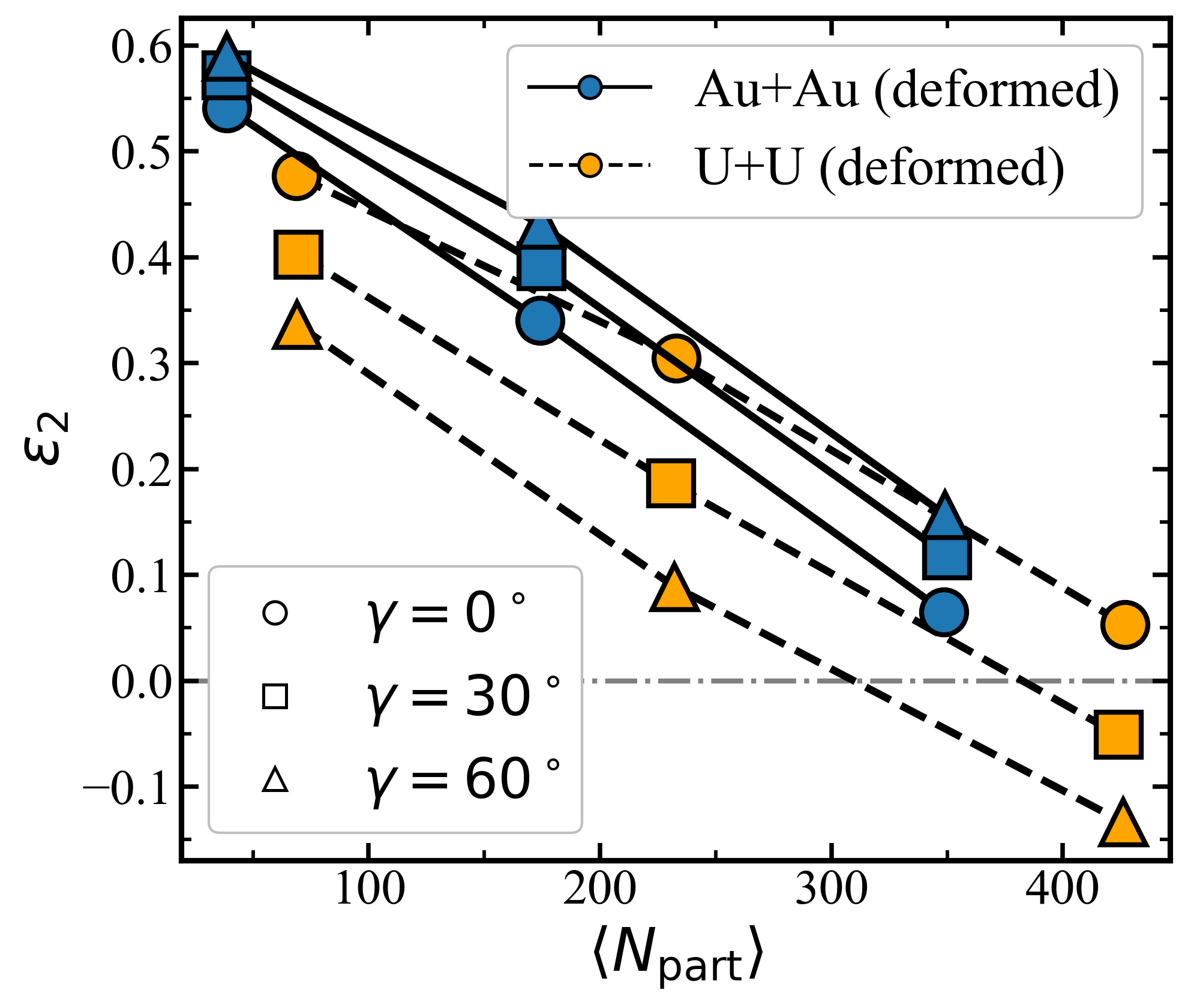}
\includegraphics[width=0.42\linewidth]{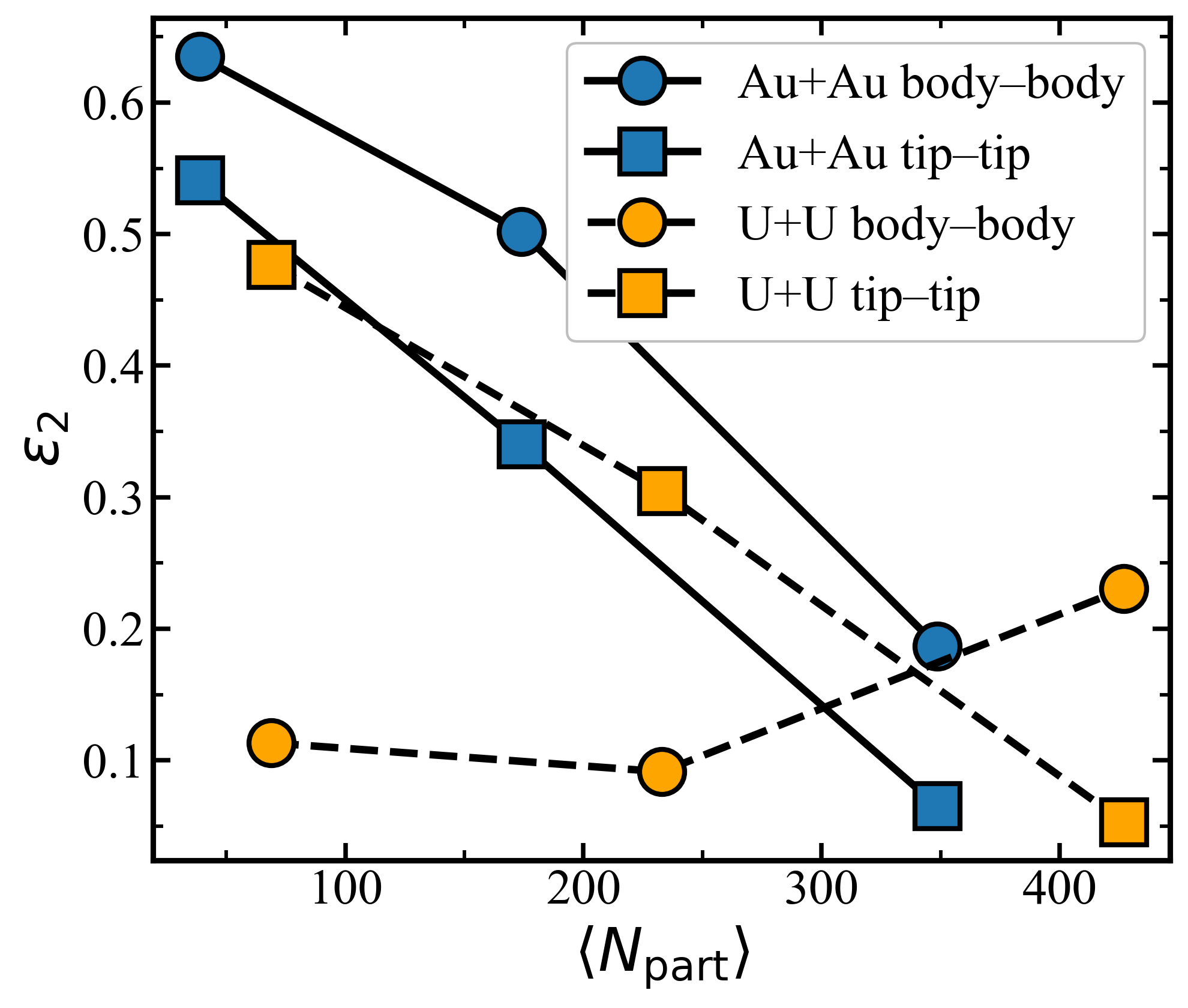}
\caption{Left panel: Elliptic eccentricity, $\varepsilon_2$, calculated using a deformed nuclear distribution characterized by the triaxiality angle $\gamma$ ($\gamma = 0^\circ, 30^\circ, 60^\circ$) for orientation-averaged Au+Au ($\sqrt{s_{NN}} = 200$ GeV) and U+U ($\sqrt{s_{NN}} = 193$ GeV) collisions. Various markers represent the triaxiality angles: circles denote $\gamma = 0^\circ$ (prolate-like), squares denote $\gamma = 30^\circ$ (triaxial), and triangles denote $\gamma = 60^\circ$ (oblate-like). Right panel: Elliptic eccentricity of the initial medium, $\varepsilon_2$, for tip–tip and body–body configurations with the triaxiality parameter $\gamma = 0^\circ$.
   }
\label{fig:glauber_moments_eps2}
\end{figure}

These variations in the anisotropic initial entropy density, stemming from the deformed nuclear density, result in distinct charmonium collective flow coefficients. We calculate the initial elliptic eccentricity of the medium entropy density $\varepsilon_2$, across various collision centralities, defined as follows,
\begin{equation}
  \varepsilon_2
  = \frac{\langle y^2 \rangle - \langle x^2 \rangle}{\langle y^2 \rangle + \langle x^2 \rangle} \,.
  \label{eq:eccentricity}
\end{equation}
where $\langle x^2\rangle$ and $\langle y^2\rangle$ characterize the spatial distribution of the medium's initial entropy density $s_n(\tau_0, \boldsymbol{x}_T)$, 
\begin{align}
  &\langle x^2 \rangle = \frac{\int d\boldsymbol{x}_T\, x^2\, s_n(\tau_0,\boldsymbol{x}_T)}{\int d\boldsymbol{x}_T s_n(\tau_0,\boldsymbol{x}_T)}\,
\end{align}
similar for $\langle y^2\rangle$. 
Fig.~\ref{fig:glauber_moments_eps2} displays the elliptic eccentricities, $\varepsilon_2$, as a function of the mean number of participating nucleons, $\langle N_{\rm part}\rangle$, for orientation-averaged Au+Au and U+U collisions. The eccentricities are calculated for triaxiality angles: $\gamma = 0^\circ, 30^\circ,$ and $60^\circ$. In U+U collisions in Fig.~\ref{fig:entropy_profiles-GAMMA}, an increase in $\gamma$ which causes the initial entropy density of the bulk medium to extend along the $x$-axis, leads to a decreases of around 0.2 in the eccentricity $\varepsilon_2$ from $\gamma=0^\circ$ to $\gamma=60^\circ$. This effect is much smaller in the case of Au+Au collisions. Such geometric modifications are expected to leave imprints on charmonium momentum anisotropies as the states evolve through the deformed hot QCD medium, primarily driven by path-length-dependent effects.

In the right panel of Fig.~\ref{fig:glauber_moments_eps2}, the body–body and tip–tip configurations in U+U collisions exhibit a pronounced separation. In peripheral collisions, the $\varepsilon_2$ in body-body collisions is significantly smaller than the case of tip-tip collisions. However, this relationship is reversed in central collisions. 
These orientation-selected U+U collisions are expected to exhibit a bit stronger geometry-driven response in charmonium momentum anisotropy, compared to the variations induced by the triaxiality angle $\gamma$ in the nuclear density.

\section{Charmonium suppression and anisotropic flow}
\label{sec:results}

\subsection{Charmonium suppression in Au+Au}

To establish a baseline prior to investigating nuclear deformation effects in U+U collisions, we first perform calculations for Au+Au collisions and validate the model against experimental data from the STAR Collaboration. Since our objective is to probe the initial geometry of the hot QCD medium arising from nuclear deformation, we focus on charmonia in high-$p_T$ bins (e.g., $p_T > 4$ GeV/c) as selective probes. These high-$p_T$ states are produced primordially, preceding the formation of the hot medium, and subsequently traverse the QGP, where they undergo varying degrees of suppression governed by the local medium temperature along their respective trajectories~\cite{Chen:2019qzx}. In contrast, regenerated charmonia emerge near the hadronization stage and exhibit diminished sensitivity to the initial geometric configuration of the medium, a consequence of the partial thermalization of the heavy-quark momentum distribution at RHIC energies.

\begin{figure}[htp]
  \centering
\includegraphics[width=0.42\linewidth]{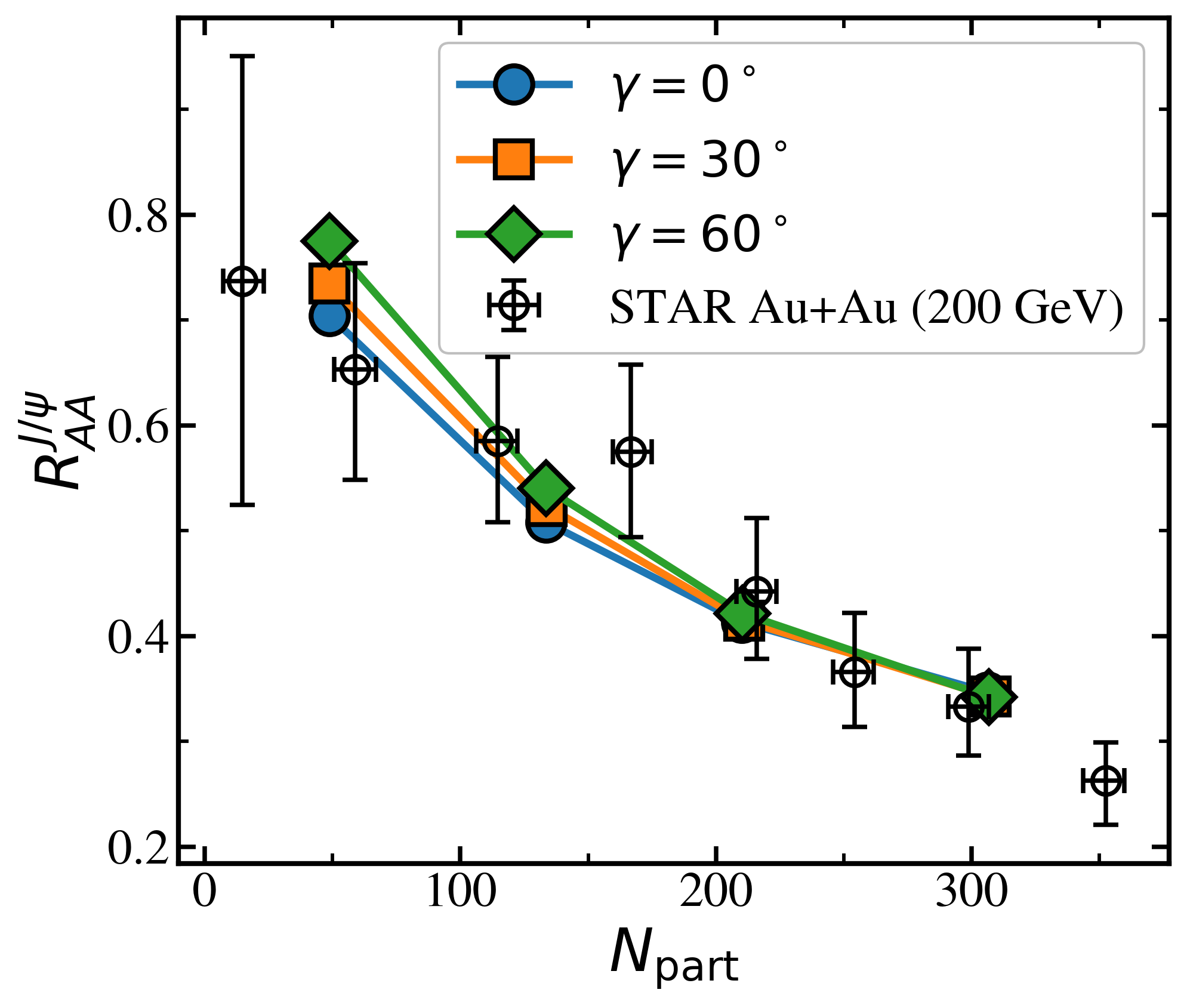}  
\includegraphics[width=0.42\linewidth]{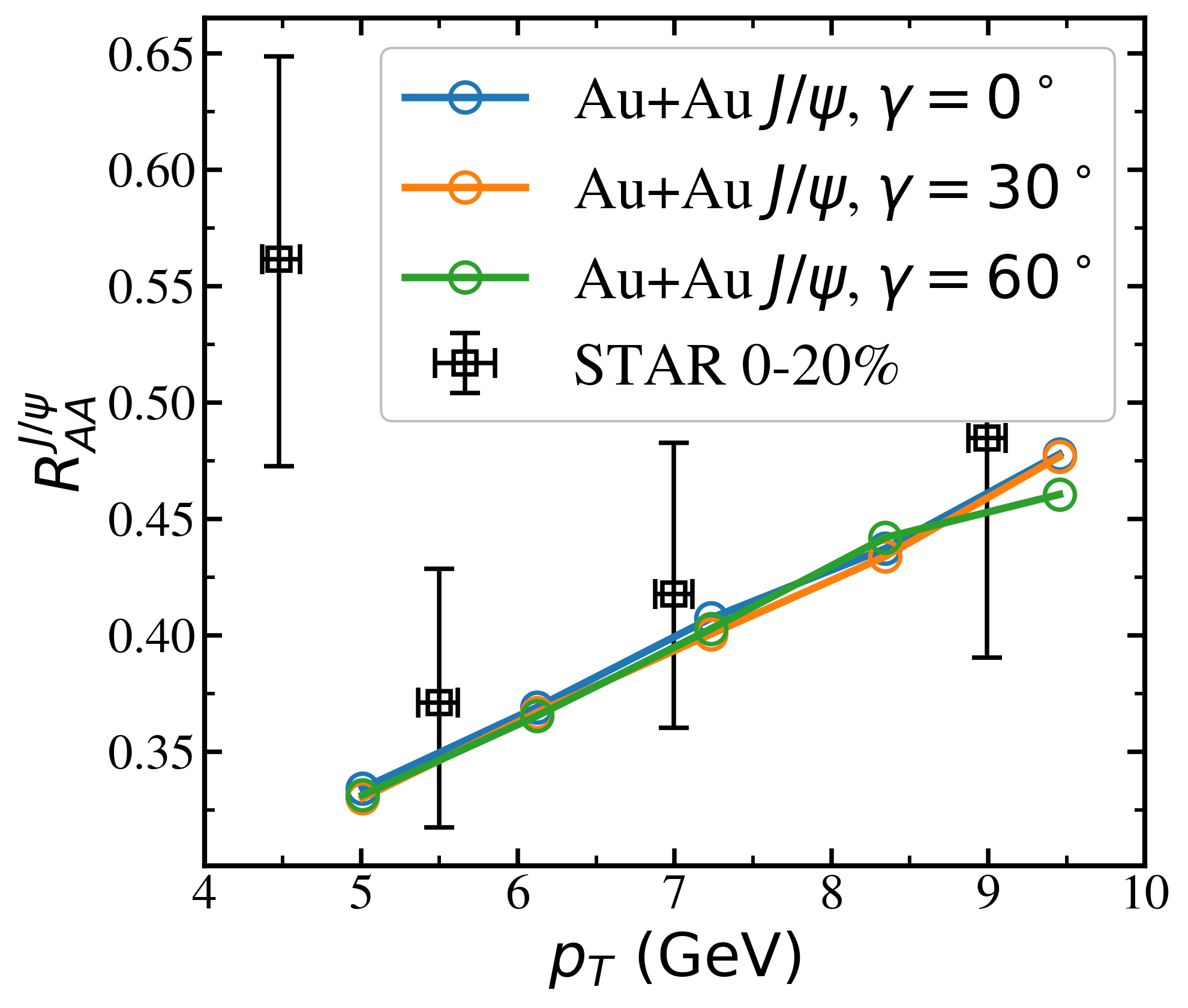}
\includegraphics[width=0.42\linewidth]{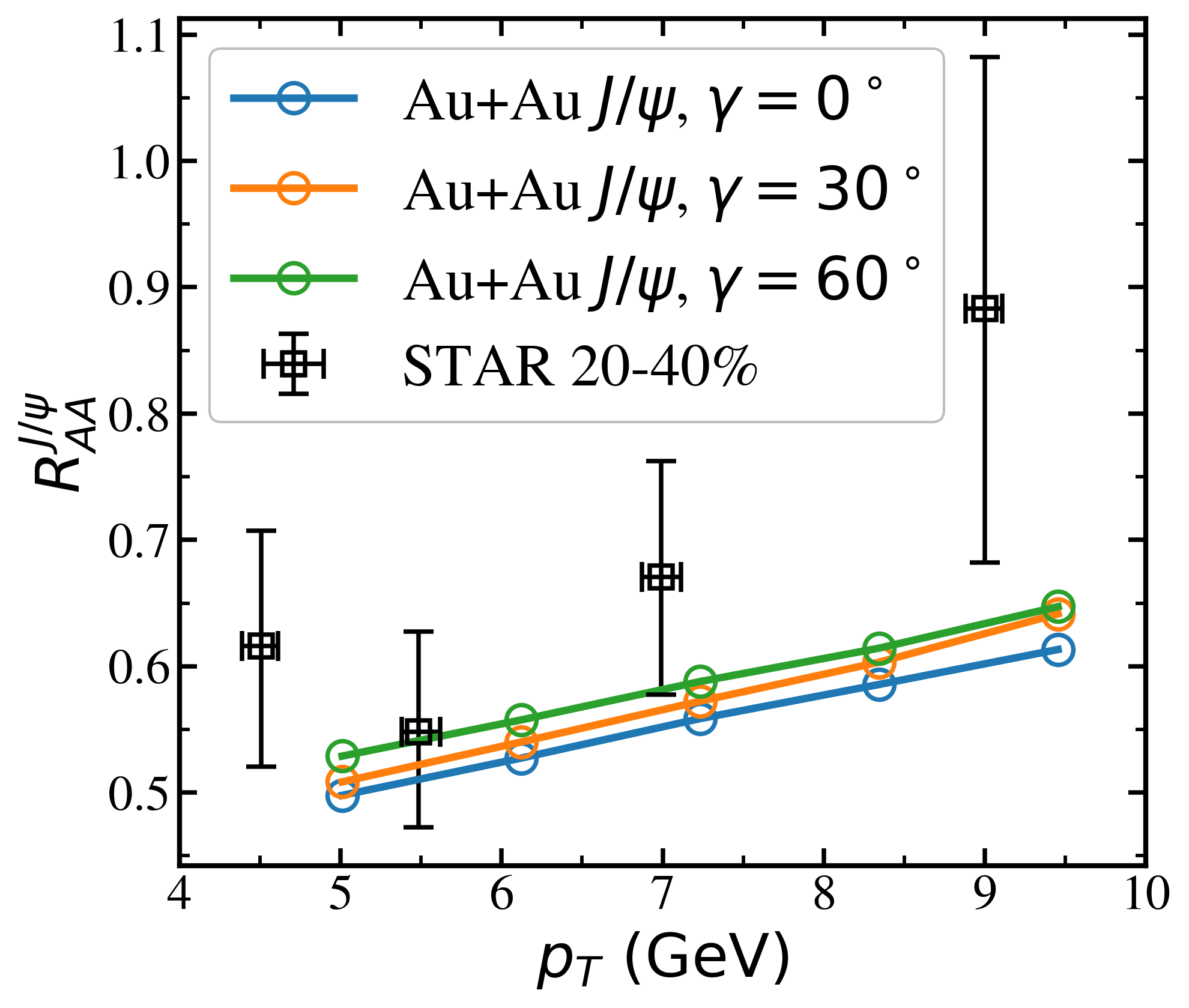}
\includegraphics[width=0.42\linewidth]{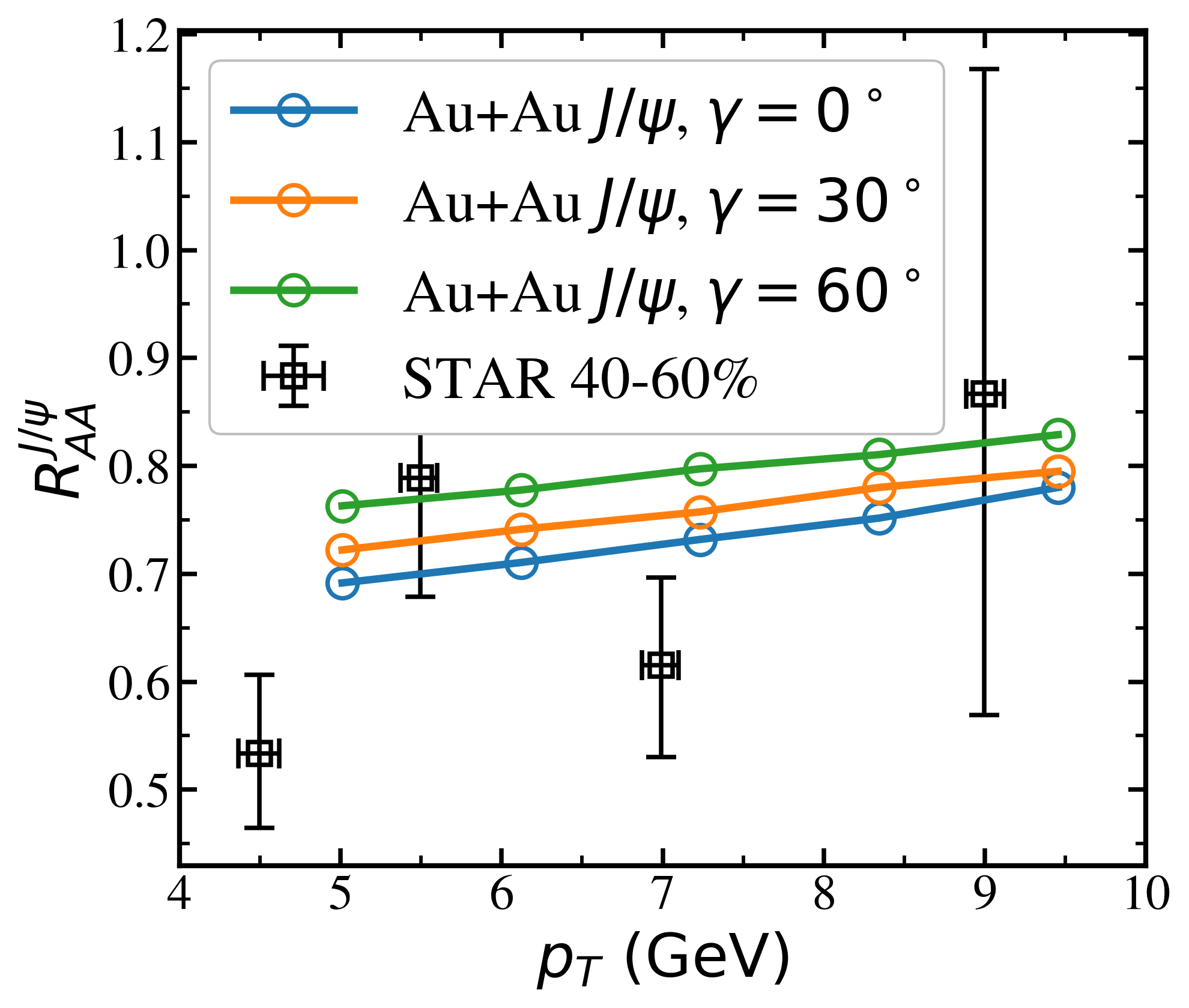}
  \caption{$J/\psi$ nuclear modification factor as a function of the number of nucleon participants within $p_T>4$ GeV/c bin, and also as a function of $p_T$ in the centrality 0-20\%, 20\%-40\%, 40\%-60\% respectively. Experimental data is from the STAR Collaboration~\cite{Adam:2019PLB}. To check the effect of nuclear deformation in Au, various values of triaxiality angle $\gamma=0^\circ,30^\circ,60^\circ$ (orientation-averaged) are considered.}
  \label{fig:RAA_pT_Np_Au}
\end{figure}

Fig.\ref{fig:RAA_pT_Np_Au} displays the $J/\psi$ nuclear modification factor, $R_{AA}$, as a function of the number of participating nucleons $N_p$, and the transverse momentum $p_T$, in 200 GeV Au+Au collisions. In the panel showing $R_{AA}(N_p)$, as $N_p$ increases, charmonium dissociation via inelastic collisions with the anisotropic medium becomes more pronounced. The theoretical calculations show good agreement with the experimental data from the STAR Collaboration~\cite{Adam:2019PLB}. To investigate the effect of the triaxiality angle $\gamma$, which introduces anisotropy into the nuclear density, various deformation configurations are tested with $\gamma = 0^\circ, 30^\circ,$ and $60^\circ$. This leads to a minor deviation in $R_{AA}$, which diminishes further in more central collisions (as indicated by the different colored solid lines in Fig. \ref{fig:RAA_pT_Np_Au}). To examine this effect across different $p_T$ intervals, $R_{AA}(p_T)$ is calculated for centrality bins of 0–20\%, 20–40\%, and 40–60\%, as shown in Fig. \ref{fig:RAA_pT_Np_Au}. Since both $R_{AA}(N_p)$ and $R_{AA}(p_T)$ result from integrations over the azimuthal angle, these observables are expected to be less sensitive to the anisotropic distribution of the bulk medium. The influence of $\gamma$ is negligible in the 0–20\% centrality class but becomes more discernible in the 40–60\% bin. The impact of varying $\gamma$ on $R_{AA}(p_T)$ is consistent with the trends observed in the $R_{AA}(N_p)$ results.

\subsection{Charmonium suppression in U+U}

\begin{figure*}[htp]
  \centering
\includegraphics[width=0.29\linewidth]{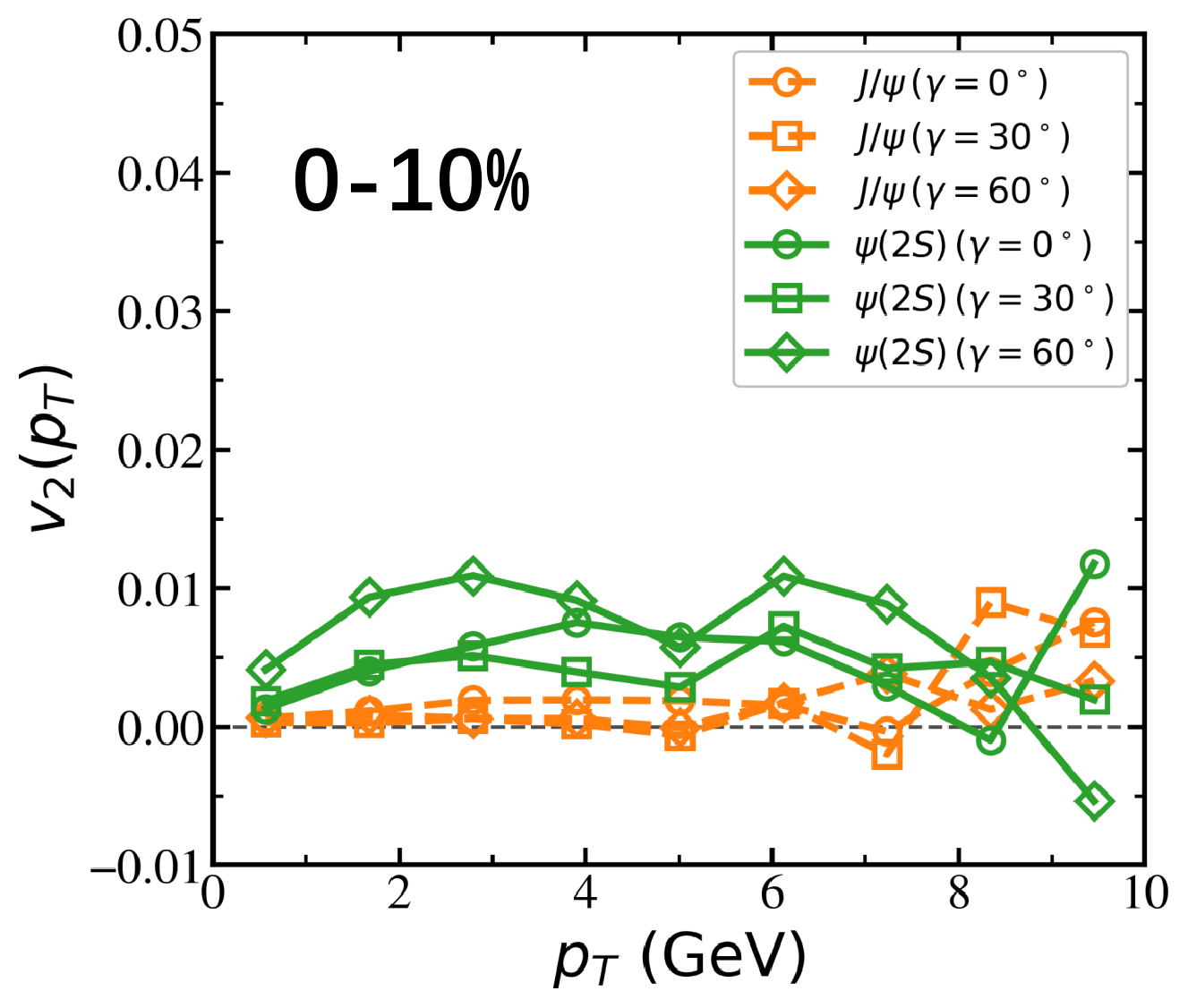}
\includegraphics[width=0.29\linewidth]{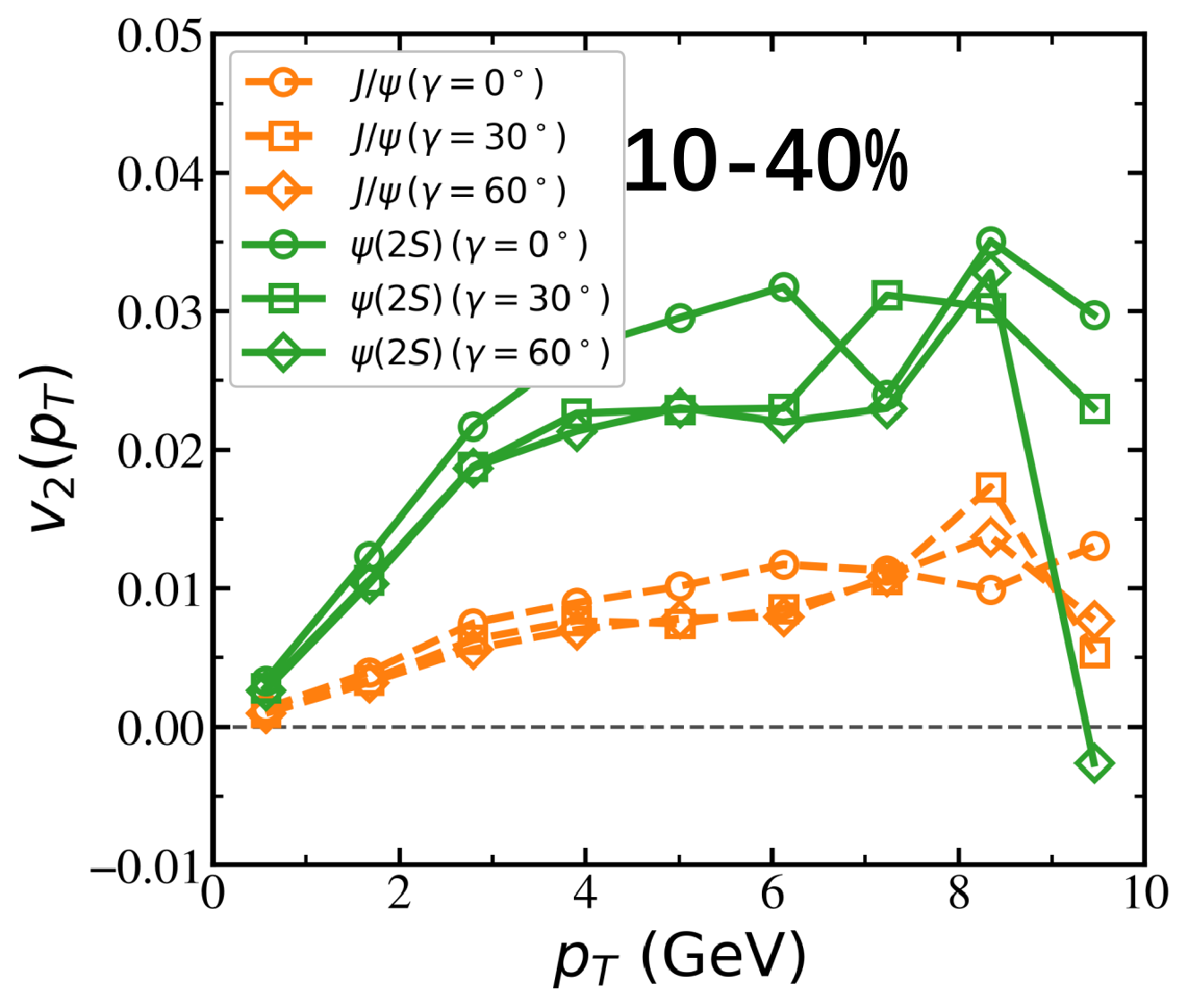}
\includegraphics[width=0.29\linewidth]{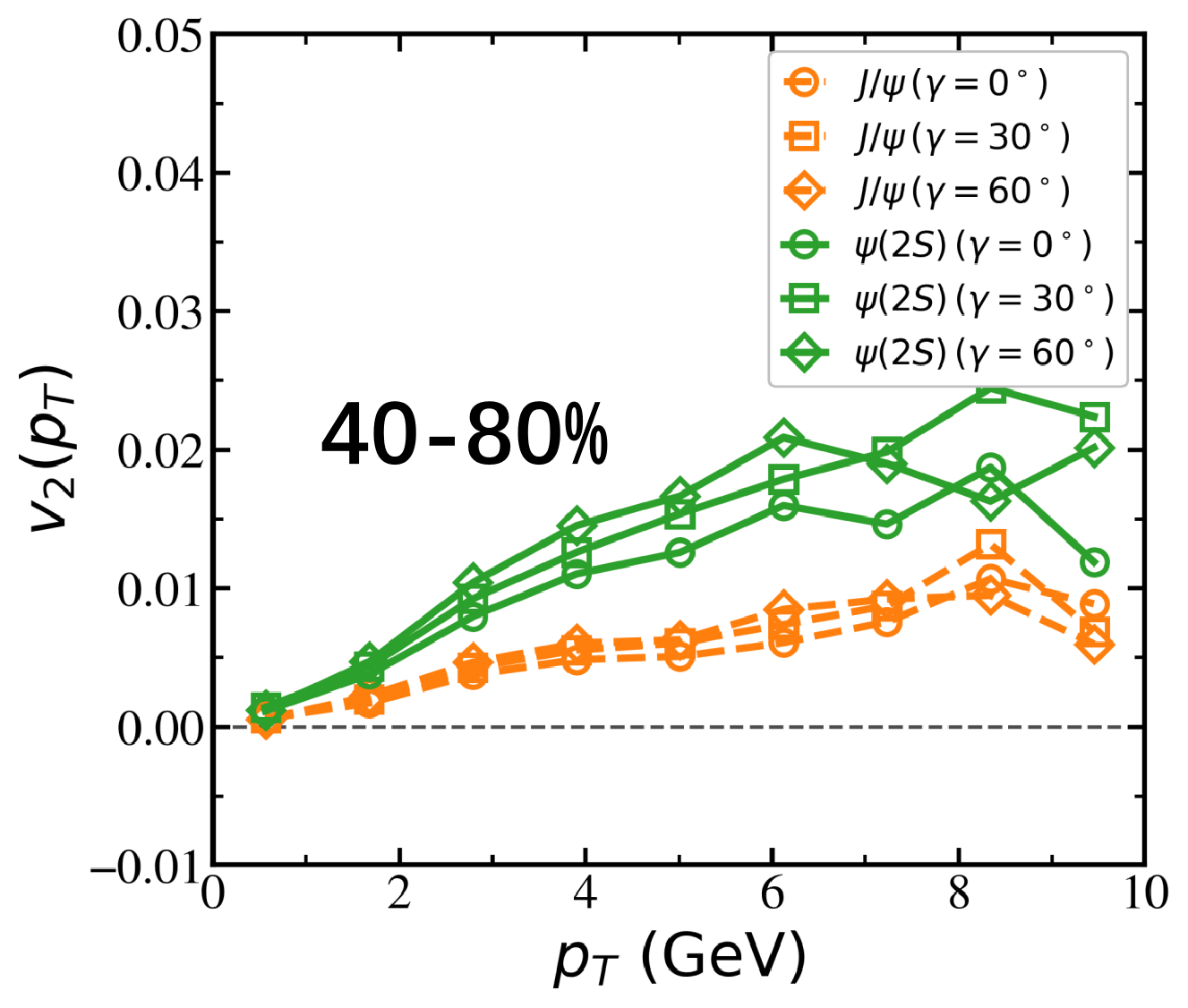}
  \caption{Orientation-averaged elliptic flow coefficients of $J/\psi$ (dashed lines) and $\psi(2S)$ (solid lines) as a function of $p_T$ in various centralities in $\sqrt{s_{NN}}=193$ GeV U+U collisions. The parameter $\gamma$ characterizing the nuclear deformation is taken as $\gamma=0^\circ,30^\circ,60^\circ$.}
  \label{fig:RAA_v2_pT_U}
\end{figure*}

In U+U collisions, the initial conditions for both charmonia in the transport model and the bulk medium in the hydrodynamic model are updated accordingly. Fig.\ref{fig:RAA_v2_pT_U} presents the orientation-averaged elliptic flow coefficient, $v_2$, for $J/\psi$ and $\psi(2S)$ across various Bohr triaxiality angles $\gamma$. Only the primordial production of charmonia is considered. In various centralities, the $v_2$ of primordially produced $J/\psi$ exhibits relatively minor sensitivity to the triaxiality values ($\gamma = 0^\circ, 30^\circ, 60^\circ$), as shown by the dashed lines with different markers. This elliptic flow is primarily attributed to the path-length-difference effect, wherein charmonia experience varying degrees of suppression along distinct trajectories. Regarding the excited state $\psi(2S)$, its binding energy is significantly lower than that of the ground state, rendering it more sensitive to the spatial anisotropy of the medium's energy density. Consequently, the $v_2$ of $\psi(2S)$ is larger than that of the $J/\psi$ in the 10–40\% and 40–80\% centrality bins, see the solid lines in Fig.\ref{fig:RAA_v2_pT_U}. Furthermore, the impact of nuclear deformation is slightly larger in the $\psi(2S)$ $v_2$ within these centralities. It is worth noting a non-linear relationship between the nuclear deformation parameter $\gamma$ and charmonium elliptic flow across different centralities. The $v_2$ magnitude is determined by both the intrinsic nuclear shape (governed by $\gamma$) and the geometric overlap area of the colliding nuclei controlled by the impact parameter $b$, a feature clearly visible in the 10–40\% and 40–80\% samples.

\begin{figure*}[htp]
  \centering
\includegraphics[width=0.3\linewidth]{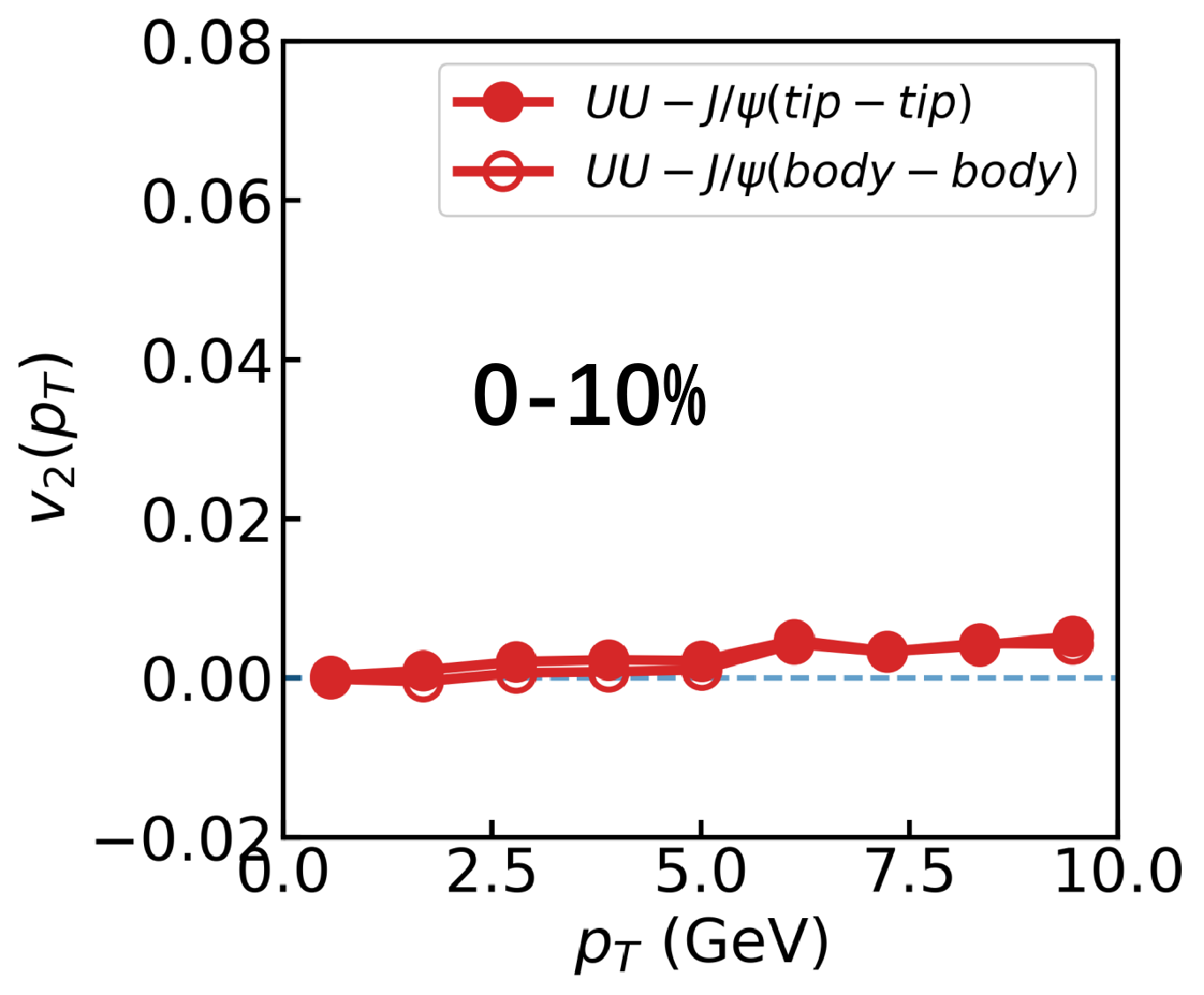}
\includegraphics[width=0.3\linewidth]{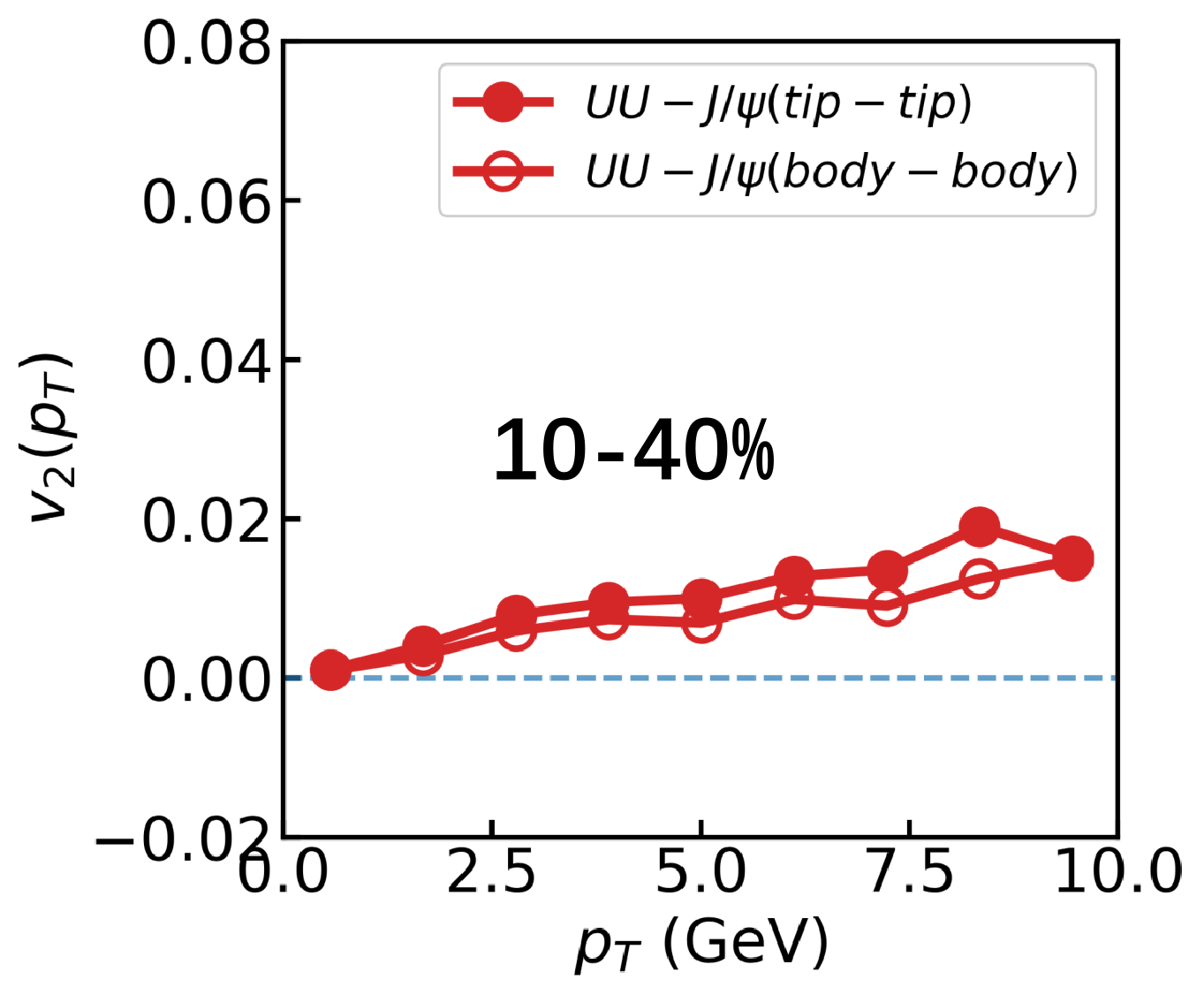}
\includegraphics[width=0.3\linewidth]{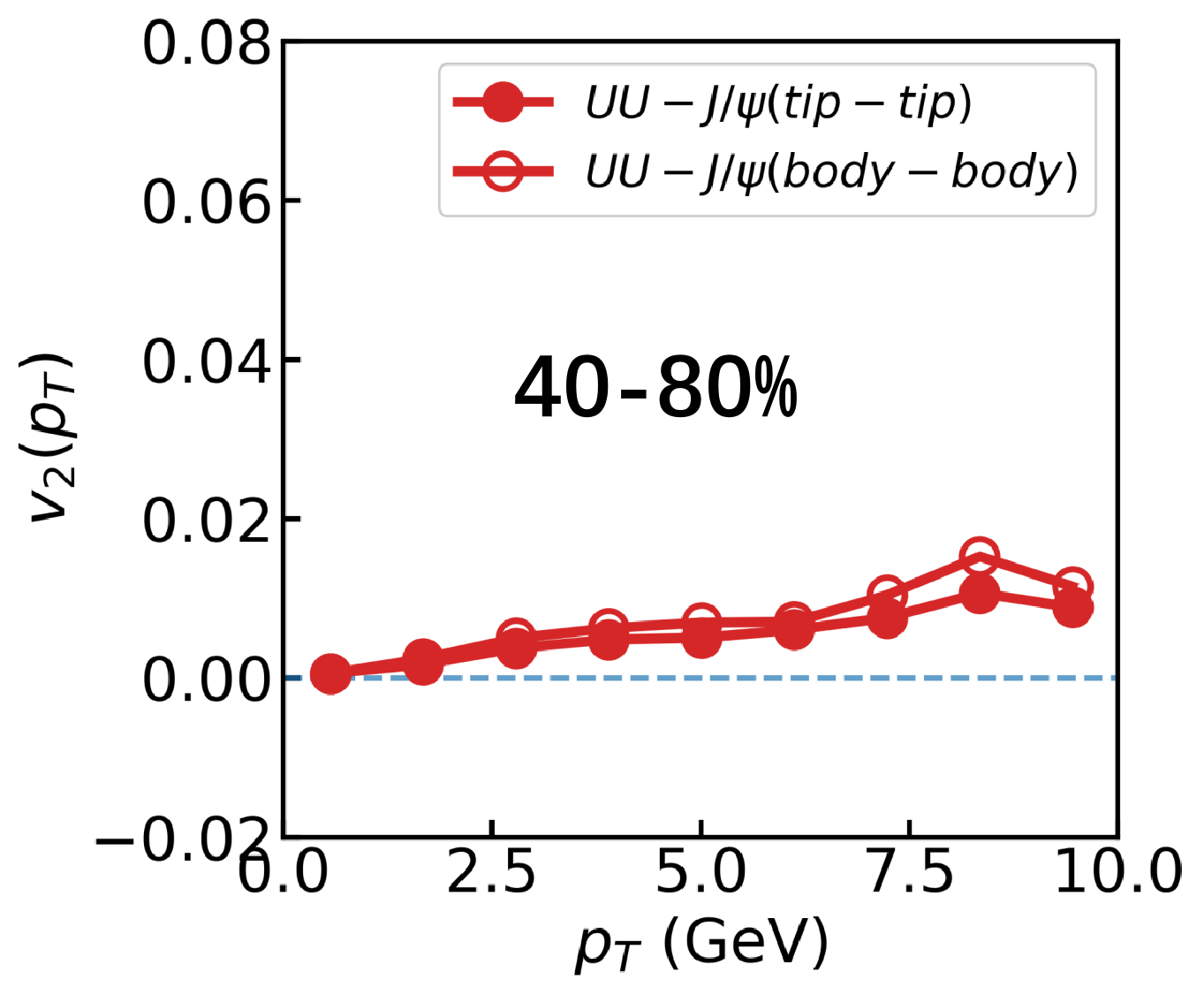}
\includegraphics[width=0.3\linewidth]{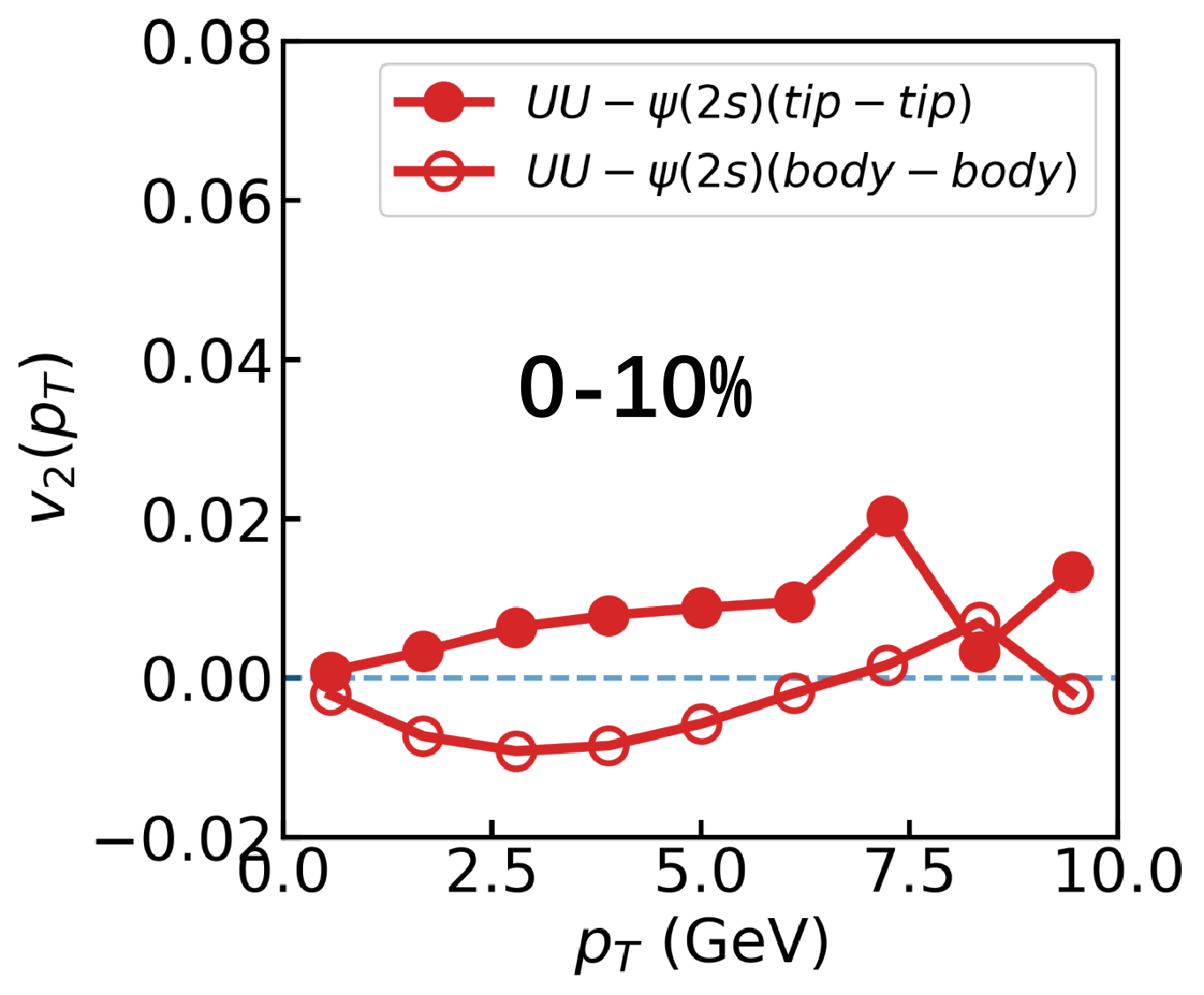}
\includegraphics[width=0.3\linewidth]{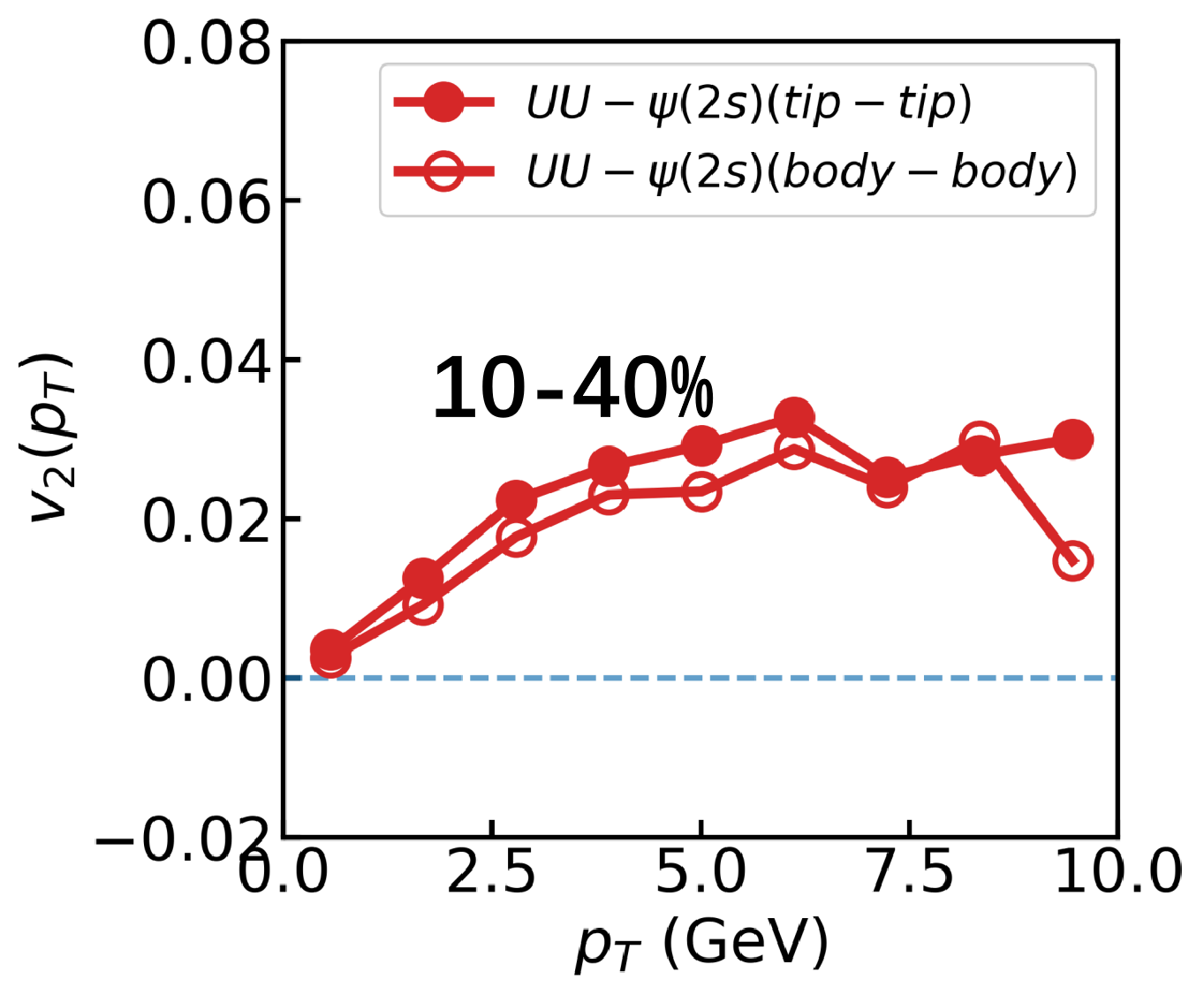}
\includegraphics[width=0.3\linewidth]{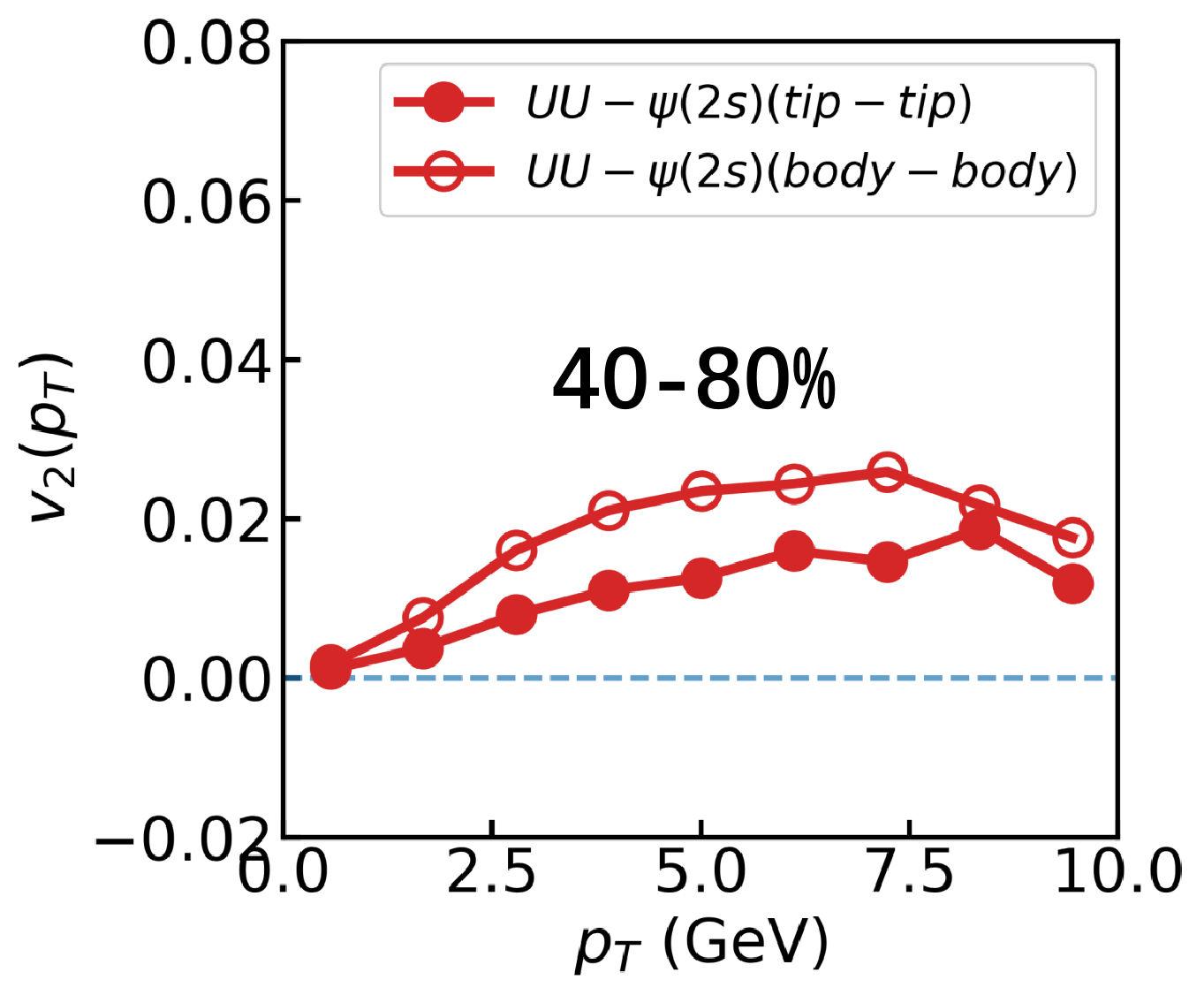}
  \caption{Elliptic flow coefficient $v_2$ of $J/\psi$ and $\psi(2S)$ in $\sqrt{s_{NN}}=193$~GeV for idealized tip-tip and body-body collisions at $\gamma=0^\circ$. Panels show the centrality classes 0--10\%,
  10--40\%, and 40--80\%.}
  \label{fig:Jpsi_tiptip_body_v2}
\end{figure*}

Owing to the intrinsic deformation of the Uranium nucleus, $\sqrt{s_{NN}}=193$ GeV U+U collisions exhibit distinct orientation-dependent configurations, specifically tip–tip and body–body collisions. These orientations contribute to the azimuthal anisotropy of the bulk medium's initial energy density (shown as Fig.\ref{fig:entropy_profiles-body}), which is subsequently reflected in the charmonium elliptic flow. Fig.\ref{fig:Jpsi_tiptip_body_v2} displays the calculated $v_2$ for $J/\psi$ and $\psi(2S)$ in both tip–tip and body–body U+U collisions across various centralities. For both states, a divergence in $v_2$ is observed between the tip–tip and body–body configurations, spanning from central to peripheral collisions, indicated by the circles and triangles respectively in Fig. \ref{fig:Jpsi_tiptip_body_v2}.

\section{Summary}
\label{sec:summary}

We investigated the effects of intrinsic nuclear deformation and collision orientation on charmonium observables at high transverse momentum. The nuclear density was parameterized using an anisotropic optical Glauber model, while the subsequent dynamical evolutions of the bulk medium and heavy quarkonia are described by a hydrodynamic model and a Boltzmann-type transport model, respectively. Our theoretical framework successfully reproduces charmonium observables in 200 GeV Au+Au collisions, establishing a robust baseline for calculations in 193 GeV U+U collisions. To quantify the impact of Uranium deformation, we examined various triaxiality angles $\gamma$, which introduce azimuthal anisotropy into the nuclear density distribution. This effect is relatively modest for $J/\psi$ elliptic flow $v_2$ but becomes a bit pronounced for the $\psi(2S)$ state, owing to its lower binding energy and heightened sensitivity to the medium's spatial distribution. Furthermore, we calculated the $v_2$ for idealized tip–tip and body–body U+U configurations. A small disparity in $v_2$ is observed between these two orientations across various centralities. This study help to explore the nuclear deformation on the aspect of heavy quarkonium.

\vspace{2cm}
\begin{acknowledgments}
This work is supported by the
National Natural Science Foundation of China (NSFC)
under Grant Nos. 12575149 and 12175165.
\end{acknowledgments}

\bibliographystyle{apsrev4-1}
\bibliography{refs}

\end{document}